\documentclass[floatfix,aps,prd,preprintnumbers,nofootinbib,superscriptaddress,preprint]{revtex4-1}

\usepackage{amsmath}
\usepackage{amssymb}
\usepackage{slashed}
\usepackage{mathtools}
\usepackage[utf8]{inputenc}
\usepackage[colorlinks=true,linkcolor=blue,citecolor=blue,urlcolor=blue,bookmarksopen,bookmarksnumbered]{hyperref}
\usepackage{color}
\usepackage[usenames,dvipsnames]{xcolor}
\usepackage[normalem]{ulem}
\usepackage{soul}
\usepackage{units}
\usepackage{rotating}
\usepackage{hhline,multirow,tabularx}
\usepackage{subfigure}
\usepackage{commath}
\usepackage{graphicx,bm}
\usepackage{verbatim}

\newcommand{\no}{\nonumber\\}
\newcommand{\parz}[1]{\ensuremath{\left(#1\right)}}

\newcommand{\xbj}{\ensuremath{x_{\mbox{\tiny \rm Bj}}}}
\newcommand{\xF}{\ensuremath{x_{\mbox{\tiny \rm F}}}}

\newcommand{\diff}[1]{\mathrm{d}#1}

\newcommand{\eref}[1]{Eq.~(\ref{e.#1})}

\newcommand{\fref}[1]{Fig.~\ref{f.#1}}

\newcommand{\aref}[1]{Appendix~\ref{a.#1}}
\newcommand{\sref}[1]{Sec.~\ref{s.#1}}
\newcommand{\ssref}[1]{Section~\ref{ss.#1}}

\newcommand{\msbar}{\ensuremath{\overline{\rm MS}}}

\begin{document}

\title{Simultaneous Monte Carlo analysis of parton densities \\
and fragmentation functions}

\preprint{JLAB-THY-21-3304}

\author{E.~Moffat}
\email{emoff003@odu.edu}
\affiliation{Department of Physics, Old Dominion University, Norfolk, Virginia 23529, USA}
\author{W.~Melnitchouk}
\email{wmelnitc@jlab.org}
\affiliation{Jefferson Lab,
	     Newport News, Virginia 23606, USA \\
        \vspace*{0.2cm}
        {\bf Jefferson Lab Angular Momentum (JAM) Collaboration
        \vspace*{0.2cm} }}
\author{T.~C.~Rogers}
\thanks{trogers@odu.edu - \href{https://orcid.org/0000-0002-0762-0275}{ORCID: 0000-0002-0762-0275}} 
\affiliation{Department of Physics, Old Dominion University, Norfolk, Virginia 23529, USA}
\affiliation{Jefferson Lab,
	     Newport News, Virginia 23606, USA \\
        \vspace*{0.2cm}
        {\bf Jefferson Lab Angular Momentum (JAM) Collaboration
        \vspace*{0.2cm} }}
\author{N.~Sato}
\email{nsato@jlab.org}
\affiliation{Jefferson Lab,
	     Newport News, Virginia 23606, USA \\
        \vspace*{0.2cm}
        {\bf Jefferson Lab Angular Momentum (JAM) Collaboration
        \vspace*{0.2cm} }}

\date{\today}

\begin{abstract}
We perform a comprehensive new Monte Carlo analysis of high-energy lepton-lepton, lepton-hadron and hadron-hadron scattering data to simultaneously determine parton distribution functions (PDFs) in the proton and parton to hadron fragmentation functions (FFs).
The analysis includes all available semi-inclusive deep-inelastic scattering and single-inclusive $e^+ e^-$ annihilation data for pions, kaons and unidentified charged hadrons, which allows the flavor dependence of the fragmentation functions to be constrained.
Employing a new multi-step fitting strategy and more flexible parametrizations for both PDFs and FFs, we assess the impact of different data sets on sea quark densities, and confirm the previously observed suppression of the strange quark distribution.
The new fit, which we refer to as ``JAM20-SIDIS'', will allow for improved studies of universality of parton correlation functions, including transverse momentum dependent (TMD) distributions, across a wide variety of process, and the matching of collinear to TMD factorization descriptions.
\end{abstract}

\maketitle

\section{Introduction}
\label{s.intro}

The standard parton correlation functions of QCD, such as collinear parton distribution functions (PDFs) and fragmentation functions (FFs), are being utilized in an increasingly diverse range of phenomenological applications.
Beyond their traditional role in predicting new high energy phenomena, they also enter frequently into the study of more complex and extended objects like transverse momentum dependent (TMD) PDFs and FFs and generalized parton distributions (GPDs), where they are needed to understand the transition between different factorization regions. 
Both TMDs and GPDs are central to the study of the nonperturbative parton structure of hadrons, and understanding how they encapsulate their longitudinal and transverse features will be critical to current experimental programs at Jefferson Lab and elsewhere, as well as to the future Electron-Ion Collider.
These considerations provide one of the main motivations for the study of collinear PDFs and FFs in this paper.

The great value of PDFs and FFs extracted from global QCD data analysis lies with their predictive power, or ``universality''. 
However, the translation from experimental data to quark and gluon operator structures is a challenging inverse problem.
It is not possible to exactly constrain parton correlations from data alone since this connection involves nontrivial convolution integrals in a factorization formalism (whose accuracy itself is difficult to quantify in any given instance), and because of the limited quantity of available data.
The complexity of the inverse problem is also magnified by the number of flavor degrees of freedom involved.

Nevertheless, assessing and maximizing the universality of collinear PDFs and FFs is crucial given the increasingly broad scenarios where they are used.  
A major focus in the current effort by the Jefferson Lab Angular Momentum (JAM) Collaboration is therefore to both test and broaden the predictive power of parton correlation functions. 
This is achieved through a Bayesian inference procedure in which PDFs and FFs are extracted simultaneously, and the uncertainty quantification associated with particular parametrizations of parton correlation functions is given in terms of a Bayesian posterior distribution. 
To test universality, the system of equations relating observables to parton correlation functions must of course exceed the total number of correlation functions involved --- a minimum requirement is that the parton correlation functions be overconstrained by the data in the fit.
Of course, realizing this in practical analyses requires that all parton correlation functions be truly fitted simultaneously.
This is a major numerical and technological challenge, and traditionally PDFs and FFs have thus been extracted in separate procedures.
However, simultaneous fits can be achieved with the Bayesian Monte Carlo approach, and have been implemented recently in the JAM17~\cite{Ethier:2017zbq} analysis of helicity PDFs, and in the JAM19~\cite{Sato:2019yez} analysis of unpolarized PDFs and FFs.
The same basic methodology was also applied in the three-dimensional JAM3D20~\cite{Cammarota:2020qcw} study, in the first combined analysis of TMD observables that satisfies the overconstraining criterion.

In this paper, we extend the previous work by performing the first simultaneous and overconstrained fit of unpolarized PDFs and FFs that utilizes both charged hadron production in semi-inclusive deep-inelastic scattering (SIDIS) and single-inclusive $e^+ e^-$ annihilation (SIA). 
This is partly motivated by a number of recent observations associated with the study of TMD PDFs.
For example, significant tension has recently been found between fits performed with standard sets of PDFs and FFs and fixed order perturbative QCD calculations in processes including SIDIS~\cite{Gonzalez-Hernandez:2018ipj, Wang:2019bvb}, Drell-Yan (DY)~\cite{Bacchetta:2019tcu}, and SIA into wide-angle hadron pairs~\cite{Moffat:2019pci}.
A number of suggested solutions and explanations have been proposed to account for this, including a possible need for power suppressed corrections~\cite{Liu:2019srj} at the moderate scales of most SIDIS experiments. 
However, more tests of the limits of applicability of standard collinear factorization are needed before it is possible to draw firm conclusions.
Given that the majority of data used to constrain collinear correlation functions (both PDFs and FFs) are either highly inclusive or exist are at very high scales, or both, it is perhaps not surprising that tension arises when these are evolved downward and used to make predictions at lower scales and for highly differential observables.
Indeed, there have been few tests that $Q^2$-scaling, a hallmark of the collinear perturbative regime, actually holds to a reasonable approximation in SIDIS measurements at moderate $Q^2$.
Our hope is that the new combined fit, which we refer to as ``JAM20-SIDIS'', will help to shed light on this and similar issues in the future.

In Sec.~\ref{s.method} we begin the discussion by summarizing the methodology used in our simultaneous Monte Carlo analysis, including the parametrizations used for the distributions and the multi-step Bayesian inference algorithm.
Details of the data sets included in the fit are summarized in \sref{datasets}, while in \sref{uni} we discuss the criteria for universality and how these are met in this analysis.
A detailed discussion of the numerical results is given in \sref{analysis}, where we present the fitted PDFs and FFs, as well as detailed comparisons of data to theory.
Finally, in \sref{conclusion} we summarize our conclusions and discuss the implications of our analysis.
Some formulas for SIDIS cross sections and structure functions are collected in \aref{sidisrev}.

\section{Theoretical framework}
\label{s.method}

In this section we give an overview of the theoretical framework on which our analysis is based, including the observables to be fitted, the parametrizations used for the PDFs and FFs, details of the perturbative QCD setup, and Bayesian inference strategy employed.

\subsection{Observables and factorization}

In this analysis we work in standard collinear factorization~\cite{Collins:1989gx, Ellis:1991qj, Collins:2011qcdbook}, in which QCD cross sections are separated into perturbatively calculable partonic hard factors convoluted with nonperturbative PDFs and/or FFs. 
We perform calculations of all observables consistently to order $\alpha_s$ in the QCD coupling.
Details of the basic theoretical setups for the inclusive DIS, inclusive Drell-Yan lepton-pair production and SIA reactions are provided in the literature~\cite{Ellis:1991qj, Devenish:2004pb}, and will not be repeated here. 
However, since SIDIS is a comparatively novel addition to global QCD analyses, we review it in more detail in \aref{sidisrev}.

The processes considered in the present analysis can be summarized as follows:
\begin{align}
\ell + N &{} \to \ell + X ,
\qquad &{}\text{inclusive DIS} \, 
\no
\ell + N &{} \to \ell + h^\pm + X ,
\qquad &{}\text{semi-inclusive DIS} \,
\no
N_1 + N_2 &{} \to \ell^+ + \ell^- + X ,
\qquad &{} \text{Drell-Yan lepton-pair production}\,  
\no
\ell^+ + \ell^- &{} \to h^\pm + X ,
\qquad &{}\text{single-inclusive annihilation}\, 
\nonumber
\end{align}
where $h^\pm$ represent charged pions, kaons, or unidentified hadrons, and the nucleon $N$ (or~$N_{1,2}$) in the initial state can be either a proton or a neutron (in practice, deuteron).
Within the framework of collinear factorization, the cross sections for each of these processes can be written schematically as convolutions of hard functions and the nonperturbative parton distribution and fragmentation functions,
\begin{align}
\frac{\diff{\sigma_\text{\tiny DIS}}}{\diff{Q^2}{} \diff{\xbj}{}}
&{}= \sum_{i} 
\mathcal{H}_i^\text{\tiny DIS} \otimes f_i \, ,
\qquad &{} 
\label{e.DIS} \\
\frac{\diff{\sigma_\text{\tiny SIDIS}}{}}{\diff{Q^2}{} \diff{\xbj}{} \diff{z_h}{} }
&{}= \sum_{ij}
\mathcal{H}_{ij}^\text{\tiny SIDIS} \otimes f_i \otimes D_j^h \, ,
\qquad &{}
\label{e.SIDIS} \\
\frac{\diff{\sigma_\text{\tiny DY}}{}}{\diff{Q^2}{} \diff{\xF}{}}
&{}= \sum_{ij}
\mathcal{H}_{ij}^\text{\tiny DY} \otimes f_i \otimes f_j\, , 
\qquad &{}
\label{e.DY} \\
\frac{\diff{\sigma_\text{\tiny SIA}}{}}{\diff{Q^2}{} \diff{z_h}{}}
&{}= \sum_{j}
\mathcal{H}_j^\text{\tiny SIA} \otimes D_j^h\, ,
\qquad &{}
\label{e.SIA}
\end{align}
where the symbols $\otimes$ represent the convolution integrals in longitudinal momentum fractions of the hard scattering functions $\mathcal{H}_{ij}$ and the PDFs $f_i$ and FFs $D_j^h$ for parton flavors $i, j$ (see the Appendix).
In each process, $Q$ represents the hard scale given by the photon virtuality, $Q \gg$ hadron masses, which allows the observables to be factorized into the short-distance perturbative and long-distance nonperturbative parts.

For the inclusive DIS and SIDIS processes,
\begin{equation}
\xbj = \frac{Q^2}{2 p \cdot q}
\end{equation}
is the usual Bjorken scaling variable, while for the DY process the analogous scaling variables are defined as
\begin{equation}
x_1 = \frac{Q^2}{2 p_1 \cdot q}\, , \qquad
x_2 = \frac{Q^2}{2 p_2 \cdot q}\, ,
\end{equation}
where $p_1$ and $p_2$ denote the incoming hadron momenta, with the Feynman scaling variable given by
\begin{equation}
\xF = x_1 - x_2.
\end{equation}
In the DY center of mass frame, and in the limit of negligible hadron masses ($\ll Q$), the virtual photon rapidity can be written in terms of $x_1$ and $x_2$ as 
\begin{equation}
y = \frac12 \ln\frac{x_1}{x_2}.
\end{equation}
For the processes involving fragmentation to a hadron $h$ in the final state, we have
\begin{equation}
\hspace*{4cm}
z_h = \frac{p_h \cdot p}{q \cdot p}
\hspace*{4cm} \textrm{[SIDIS]}
\end{equation}
for SIDIS in \eref{SIDIS}, while
\begin{equation}
\hspace*{3.6cm}
z_h = \frac{2 p_h \cdot q}{Q^2}
\hspace*{3.8cm} \textrm{[SIA]}
\end{equation}
for SIA in \eref{SIA}.

\subsection{Perturbative QCD and numerical setups}

For our numerical analysis we make use of Mellin space techniques to enable fast evaluations of observables needed for the Bayesian analysis.
In particular, we solve the DGLAP evolution equations analytically in Mellin space~\cite{Vogt:2004ns}, which allows one to effectively render high-dimensional momentum space convolutions from process-specific factorization theorems, along with the integrals in the DGLAP equations, in the form of lower-dimensional inverse Mellin transforms.
For example, for the inclusive DIS observables one can write schematically 
\begin{align}
    \frac{\diff \sigma_\text{\tiny DIS}}{\diff Q^2 \diff \xbj}
    = \sum_{i}
      \frac{1}{2\pi i} \int dN\, \xbj^{-N}\,
       \widetilde{\cal H}^\text{\tiny DIS}_i(N,\mu)\,
       U^{\rm S}_{ij}(N,\mu,\mu_0)\,
       \widetilde{f}_j(N,\mu_0),
\end{align}
where $N$ here is the conjugate variable to $\xbj$, $\widetilde{f}_j(N,\mu_0)$ is the Mellin moment of the PDF $f_j(x,\mu_0)$, defined by
\begin{equation}
\label{e.fNdef}
\widetilde{f}_j(N,\mu_0)
= \int_0^1 dx\, x^{N-1}\, f_j(x,\mu_0),
\end{equation}
and $\widetilde{\cal H}^\text{\tiny DIS}_i(N,\mu)$ is the corresponding moment of the partonic DIS cross section.
The analytic solution for the DGLAP evolution is entirely encoded in the evolution matrix $U^{\rm S}_{i,j}$ that evolves the moments $\widetilde{f}_j(N,\mu_0)$ of the PDFs from a given input scale $\mu_0$ to the relevant DIS hard scale $\mu=Q$.
A similar expression can be written for the SIA cross section,
\begin{align}
    \frac{\diff \sigma_\text{\tiny SIA}}{\diff Q^2 \diff z_h}
    = \sum_{ij}
      \frac{1}{2\pi i} \int dM\, z_h^{-M}\,
      \widetilde{\cal H}^\text{\tiny SIA}_i(M,\mu)\,
      U^{\rm T}_{ij}(M,\mu,\mu_0)\,
      \widetilde{D}_j^h(M,\mu_0)
\end{align}
where $M$ is the Mellin conjugate variable for $z_h$, $\widetilde{D}_j^h(M,\mu_0)$ is the moment of the FF, and $\widetilde{\cal H}^\text{\tiny SIA}_i$ is the moment of the partonic SIA cross section.
The superscripts $\rm S$ and $\rm T$ in the evolution matrix  distinguish between the spacelike and timelike evolution for the PDFs and FFs, respectively, which are encoded in the corresponding DGLAP splitting kernels.

The same procedure can be extended for the case of SIDIS, which gives
\begin{align}
    \frac{\diff \sigma_\text{\tiny SIDIS}}{\diff Q^2 \diff \xbj \diff z_h}
    = &\sum_{ijkl}
      \frac{1}{(2\pi i)^2}
      \int dN\, \xbj^{-N}
      \int dM\, z_h^{-M}\,
      \widetilde{\cal H}^\text{\tiny SIDIS}_{ik}(N,M,\mu)
      \notag\\
      \times
      &\ U^{\rm S}_{ij}(N,\mu,\mu_0)\,
      \widetilde{f}_j(\mu_0)\,
      U^{\rm T}_{kl}(M,\mu,\mu_0)\,
      \widetilde{D}_j^h(M,\mu_0).
\end{align}
For the case of the Drell-Yan process, a special treatment is required since the Mellin moments for the partonic cross sections are not known.
For this we employ the strategy developed by Stratmann and Vogelsang~\cite{Stratmann:2001pb}, where by the Mellin moments are numerically pre-calculated and used as lookup tables during the analysis.
The resulting expression can be written schematically as  
\begin{align}
    \frac{\diff \sigma_\text{\tiny DY}}{\diff Q^2 \diff \xF}
    = &\sum_{ijkl}
      \frac{1}{(2\pi i)^2} 
      \int dN\,
      \int dM\,
      \widetilde{\cal H}^\text{\tiny DY}_{ik}(N,M,\mu)
      \notag\\
      \times
      &\ U^{\rm S}_{ij}(N,\mu,\mu_0)\,
      \widetilde{f}_j(\mu_0)\,
      U^{\rm S}_{kl}(M,\mu,\mu_0)\,
      \widetilde{f}_l(\mu_0),
\end{align}
where the relevant inverse Mellin factors $x_1^{-N}$ and $x_2^{-M}$ arising from the scaling variables $x_1$ and $x_2$ for the incident nucleons $N_1$ and $N_2$, respectively, in \eref{DY} are integrated numerically with the hard scattering cross section and contained inside $\widetilde{\cal H}^\text{\tiny DY}_{ik}(N,M,\mu)$.

The analytic solutions for the evolution matrices are computed at next-to-leading logarithmic accuracy using splitting kernels up to ${\cal O}(\alpha_s^2)$ and the truncated solution for the single evolution operators (see Ref.~\cite{Vogt:2004ns} for details).
We employ the zero-mass variable flavor scheme for solving the DGLAP evolution equations, setting the input scale for the PDFs and FFs at $\mu_0=m_c$.
The numerical values for the mass thresholds are taken from the PDG values in the $\overline{\rm MS}$ scheme~\cite{Zyla:2020zbs}: 
    $m_c = 1.28$~GeV and $m_b = 4.18$~GeV.
The strong coupling is evolved numerically using the QCD beta function up to ${\cal O}(\alpha_s^2)$, using the boundary condition $\alpha_s(M_Z) = 0.118$ at the $Z$ boson mass, $M_Z=91.18$~GeV.
Finally, all the process specific hard coefficients are computed at fixed next-to-leading order in pQCD, which are available in the literature~\cite{Floratos:1981hs, Kretzer:2000yf, Stratmann:2001pb, Anastasiou:2003yy}.

\subsection{Parametrization of nonperturbative functions}
\label{ss.param}

For the nonperturbative parton distribution and fragmentation functions we use standard parametrizations that have been utilized in the literature.
Namely, for the dependence on the parton momentum fraction $x$ of the PDF $f(x)$ we use the template function
\begin{equation}
f(x,\mu_0)\ \to\ T\parz{x;\boldsymbol{a}} = {\cal M} \frac{x^\alpha\parz{1-x}^\beta\parz{1+\gamma\sqrt{x}+\delta{x}}}{\int_0^1\diff{x}\, x^{\alpha+1}\parz{1-x}^\beta\parz{1+\gamma\sqrt{x}+\delta{x}}},
\label{e.template1}
\end{equation}
where $\boldsymbol{a}=\{{\cal M},\alpha,\beta,\gamma,\delta\}$ is a vector containing the shape parameters  ($\alpha$, $\beta$, $\gamma$, and $\delta$) and a normalization coefficient (${\cal M}$) to be fitted. 
The integral in the denominator ensures that the value of the normalization coefficient ${\cal M}$ is equal to the second moment ($x$-weighted integral) of the function $T(x;\boldsymbol{a})$.
For fitting the PDFs, we assume isospin symmetry to relate the PDFs in the neutron, $f_{i/n}(x)$, to those in the proton, $f_{i/p}(x) \equiv f_i(x)$,  switching the $u \leftrightarrow d$ and $p \leftrightarrow n$ labels for the light quark flavors, and taking the PDFs for other flavors equal for the proton and neutron.

In practice, we parametrize the valence $u$ and $d$ quark distributions,
    $u_v \equiv f_u - f_{\bar u}$
and $d_v \equiv f_d - f_{\bar d}$,
directly using the template function (\eref{template1}).  The gluon distribution, $g \equiv f_g$, is also directly parametrized per \eref{template1}.
For the sea quark and antiquark distributions, we use five functions parametrized as in \eref{template1}. 
These are a flavor symmetric sea function ($S$) that dominates at very low $x$ and flavor specific functions ($q_0(\bar{q}_0)$) for the $s$, $\bar{u}$, $\bar{d}$, and $\bar{s}$ that take into account the possible nonperturbative origin of the sea. 
The distributions for $s$, $\bar{u}$, $\bar{d}$, and $\bar{s}$ are constructed from these according to:
    $q(\bar q) \equiv f_{q(\bar q)} = S + q_0(\bar q_0)$.
Note that $s$ and $\bar{s}$ are parametrized separately because their contributions to the $K^+$ and $K^-$ SIDIS cross sections differ.
We do not fit the charm and bottom PDFs, and their contributions are generated purely from the DGLAP evolution. 
In total there are 8 parametrized PDF functions being fitted.  
For the valence quark PDFs $u_v$ and $d_v$ and the nonperturbative sea components $\bar{u}_0$ and $\bar{d}_0$, we use the four shape parameters as in \eref{template1}; for all other distributions we set the $\gamma$ and $\delta$ parameters to zero.  This gives 24 free shape parameters and 8 free normalization parameters.
The number of free parameters is further reduced by valence number sum rules, which constrain the normalization parameters $\cal M$ for the $u_v$, $d_v$, and $s-\bar s$ distributions, whose lowest moments are required to be 2, 1, and zero, respectively.
The normalization for the gluon PDF is determined using the momentum sum rule.
With these constraints, there is a total of 28 free parameters for the PDFs.

For the $z$ dependence of FFs, the functional form follows a similar template,
\begin{equation}
D(z,\mu_0)\ \to\ T\parz{z;\boldsymbol{a}} = {\cal M} \frac{z^\alpha\parz{1-z}^\beta\parz{1+\gamma\sqrt{z}+\delta{z}}}
{\int_0^1\diff{z}\, z^{\alpha+1}\parz{1-z}^\beta\parz{1+\gamma\sqrt{z}+\delta{z}}},
\label{e.template2}
\end{equation}
where again the integral in the denominator ensures that the coefficient ${\cal M}$ corresponds to the second moment ($z$-weighted integral) of the function.
In addition to the fragmentation to pions and kaons studied in earlier JAM analyses of SIA and SIDIS data~\cite{Sato:2016wqj, Sato:2019yez}, here we consider also the inclusive production of unidentified charged hadrons, $h^\pm$.
Accounting for unidentified hadrons can be implemented in two ways.
First, the hadron FFs can be fit independently from those for pions and kaons, as preferred by the NNPDF Collaboration~\cite{Bertone:2018ecm}.
Alternatively, one can take advantage of existing knowledge of specified hadron FFs and add a fitted residual correction to their sum.
Such an approach was adopted by de Florian, Sassot, and Stratmann (DSS)~\cite{deFlorian:2007ekg}, for example, in which a residual correction was fitted to the sum of previously obtained pion, kaon, and proton fragmentation functions.

In our analysis we follow the latter approach, but include only the pion and kaon FFs, so that the residual term $D_i^{\rm res^+}$ parametrizes the difference between the total hadron FF $D_i^{h^+}$ and the $D_i^{\pi^+}$ and $D_i^{K^+}$ functions,
\begin{equation}
    D_i^{h^+} = D_i^{\pi^+} + D_i^{K^+} + D_i^{\rm res^+}.
\end{equation}
To reduce the total number of residual FFs being fit, we assume SU(3) flavor symmetry for light quarks and antiquarks,
\begin{subequations}
\begin{eqnarray}
    D_q^{\rm res^+}=D_u^{\rm res^+} &=& D_d^{\rm res^+} = D_s^{\rm res^+}, \ \ \ \ \ \ \
    \label{e.rprel} \\
    D_{\bar{q}}^{\rm res^+}=D_{\bar{u}}^{\rm res^+} &=& D_{\bar{d}}^{\rm res^+} = D_{\bar{s}}^{\rm res^+}, \label{e.rmrel}
\end{eqnarray}
\end{subequations}
where $D_q^{\rm res^+}$ and $D_{\bar{q}}^{\rm res^+}$ are parametrized per the template (\eref{template2}).  To allow for differentiation between the residual FFs for light quarks and antiquarks, we leave ${\cal M}$ and $\beta$ for $D_{\bar{q}}^{\rm res^+}$ as free parameters, but fix $\alpha$, $\gamma$, and $\delta$ to be the same as for $D_{q}^{\rm res^+}$.
This achieves a similar constraint on the parameters as the condition used by DSS~\cite{deFlorian:2007ekg},
    $2D_{\bar{q}}^{\rm res^+} = \parz{1-z}^{\beta'} D_{q+\bar{q}}^{\rm res^+}$.

For the pion FFs, $D_i^{\pi^+}$, we reduce the number of fitted functions by grouping the light quarks into ``favored'' (valence) and ``unfavored'' (non-valence) flavors,
\begin{subequations}
\begin{align}
    D_{\rm fav}^{\pi^+} &= D_u^{\pi^+} = D_{\bar{d}}^{\pi^+},
    \label{e.pifav} 
    \\
    D_{\rm unf}^{\pi^+} &= D_d^{\pi^+} = D_s^{\pi^+} = D_{\bar{u}}^{\pi^+} =  D_{\bar{s}}^{\pi^+},
    \label{e.piunfav}
\end{align}
\end{subequations}
where $D_{\rm fav}^{\pi^+}$ and $D_{\rm unf}^{\pi^+}$ are parametrized as in \eref{template2}. For the parameters of the kaon FFs, $D_i^{K^+}$, we equate the ``unfavored'' flavors,
\begin{equation}
    D_{\rm unf}^{K^+} = D_d^{\pi^+} = D_s^{\pi^+} = D_{\bar{u}}^{\pi^+} = D_{\bar{d}}^{\pi^+},
    \label{e.Kunfav}
\end{equation}
but leave the favored FFs $D_u^{K^+}$ and $D_{\bar{s}}^{K^+}$ independent.
Here $D_{\rm unf}^{K^+}$, $D_u^{K^+}$, and $D_{\bar{s}}^{K^+}$ are parametrized per \eref{template2}. 
For the heavier flavors, we assume the charm and bottom quark and antiquark FFs to be equivalent, $D_c^{h^+} = D_{\bar{c}}^{h^+}$ and $D_b^{h^+} = D_{\bar{b}}^{h^+}$ for $h=\pi,K,{\rm res}$, with $D_c^{h^+}$ and $D_b^{h^+}$ parametrized per \eref{template2}. 
Finally, the gluon FFs $D_g^{h^+}$ for $h=\pi,K,{\rm res}$ are also parametrized according to \eref{template2}.
We use charge conjugation symmetry to relate FFs for opposite charges by
\begin{equation}
D_q^{h^+} = D_{\bar{q}}^{h^-},
\end{equation}
where $h=\pi,K,{\rm res}$. 
This results in 5 fitted functions for pions and residual hadrons, and 6 for kaons.

At this point, there are 17 shape parameters and 5 normalization parameters for residual hadrons, 20 shape parameters, and 5 normalization parameters for pions, and 24 shape parameters and 6 normalization parameters for kaons. 
The number of shape parameters is reduced further because throughout the fitting procedure, the parameters $\gamma$ and $\delta$ for the gluon, charm, and bottom FFs are fixed at zero.
In the end there are 16 free parameters to be fitted for residual charged hadron FFs, 19 free FF pion parameters, and 24 free parameters for the kaon FFs. 
Together with the 28 PDF parameters, we have a total of 87 free parameters for the fitted functions.
In addition, there are also 42 free parameters associated with normalization of various data sets, making for a total of 129 free parameters to be fitted in the analysis.

\subsection{Bayesian inference}
\label{ss.MCfit}

Our methodology for extracting nonperturbative PDFs and FFs is based on the general premise of Bayesian inference.
Namely, we use Bayes' theorem to define a multivariate probability distribution ${\cal P}$ for the shape parameters characterizing the PDFs and FFs (the posterior) at a given input scale $\mu_0$,
\begin{align}
\label{e.Pdef}
    {\cal P}(\boldsymbol{a}|{\rm data}) \sim {\cal L} (\boldsymbol{a},{\rm data})\,
    \pi(\boldsymbol{a}),
\end{align}
where $\cal L$ is a standard Gaussian likelihood function,
\begin{align}
    {\cal L}\parz{\boldsymbol{a},{\rm data}} 
    = \exp\parz{-\frac{1}{2}\chi^2\parz{\boldsymbol{a},{\rm data}}},
\end{align}
with the $\chi^2$ function defined by
\begin{equation}
    \chi^2(\boldsymbol{a}) = \sum_{i,e}
    \bigg( \frac{d_{i,e} - \sum_k r_e^k \beta^k_{i,e} - T_{i,e}(\boldsymbol{a})/N_e}
                {\alpha_{i,e}}
    \bigg)^2
  + \sum_k \big( r^k_e \big)^2
  + \left( \frac{1-N_e}{\delta N_e} \right)^2.
    \label{e.chi2}
\end{equation}
Here, $d_{i,e}$ is the value of the $i$-th data point for the experimental dataset $e$, with $T_{i,e}$ the theoretical prediction for the data point; $\alpha_{i,e}$ is the uncorrelated systematic and statistical uncertainty for each data point added in quadrature; $\beta^k_{i,e}$ is the $k$-th source of point-to-point correlated systematic uncertainties for the $i$-th bin of dataset $e$, with $r^k_e$ the related weight; and $N_e$ and $\delta{N}_e$ are the normalization and normalization uncertainty for each data set, respectively.
In Eq.~(\ref{e.Pdef}), $\pi(\boldsymbol{a})$ is the prior distribution for the set of parameters $\boldsymbol{a}$, which is used as input for a given fit to the data.

In principle, given the Bayesian posterior distribution, one can estimate confidence regions for a generic observable ${\cal O}$ (such as a PDF or a function of PDFs or FFs) by integrating over an $d$-dimensional parameter space,
\begin{subequations}
\label{e.exp_var1}
\begin{align}
    {\rm E}[\mathcal{O}]
    &=\int\diff^d\boldsymbol{a}\, \mathcal{P}\parz{\boldsymbol{a}|{\rm data}}
    \mathcal{O}\parz{\boldsymbol{a}},
    \label{e.exp1} \\
    {\rm V}[\mathcal{O}]
    &=\int\diff^d\boldsymbol{a}\, \mathcal{P}\parz{\boldsymbol{a}|{\rm data}}
    \parz{\mathcal{O}\parz{\boldsymbol{a}}-{\rm E}[\mathcal{O}]}^2,
    \label{e.var1}
\end{align}
\end{subequations}
where E and V are the expectation value and variance of the observable ${\cal O}$, respectively.
Due to the significant numerical expense of evaluating the likelihood function, the explicit usage of Eqs.~(\ref{e.exp_var1}) is often not practical.
Instead, a more efficient option is to build Monte Carlo parameter samples $\{\boldsymbol{a}_k; k=1,\ldots,n\}$, which contain all parameters, including the $N_e$ from \eref{chi2}, that are faithfully distributed according to the posterior distribution.
These can in turn be used to evaluate the integrals in Eqs.~(\ref{e.exp_var1}) as Monte Carlo sums,
\begin{subequations}
\label{e.exp_var2}
\begin{align}
    {\rm E}[\mathcal{O}]&=\frac{1}{n}\sum_{k=1}^n\mathcal{O}\parz{\boldsymbol{a}_k},
    \label{e.exp2} \\
    {\rm V}[\mathcal{O}]&=\frac{1}{n}\sum_{k=1}^n\parz{\mathcal{O}\parz{\boldsymbol{a}_k}-{\rm E}[\mathcal{O}]}^2.
    \label{e.var2}
\end{align}
\end{subequations}

\begin{figure}[t!]
    \centering
    \includegraphics[width=\textwidth]{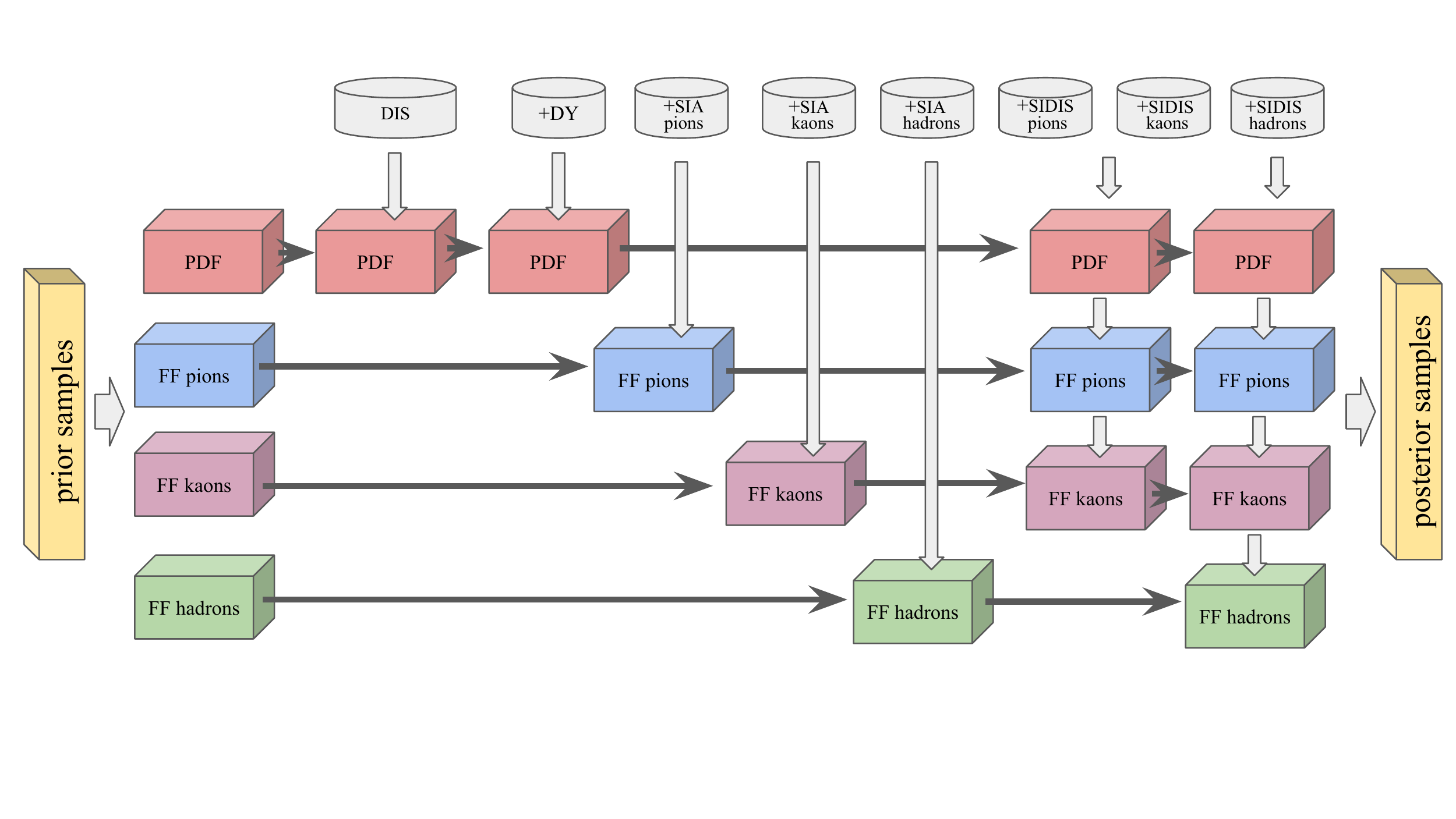}
    \vspace*{-2.5cm}
    \caption{Schematic illustration of the multi-step workflow employed in our simultaneous Monte Carlo analysis. Each box represents a collection of Monte Carlo samples associated with a specific nonperturbative hadronic structure (PDFs, FFs). The vertical arrows indicate the inclusion of additional datasets from which new optimized Monte Carlo samples (posteriors) are generated as input (priors) for the next step.\\}
    \label{f.workflow}
\end{figure}

Our Monte Carlo sampling strategy is based on data resampling methodology, whereby multiple maximum likelihood optimizations are carried out.
Each optimization consists of taking a random point in parameter space and fitting the parameters to data that have been distorted away from the central values by Gaussian shifts within the quoted uncertainties. 
To build the Monte Carlo samples, we use the multi-step strategy developed in Ref.~\cite{Sato:2019yez}, where the PDF and FF parameters are pre-optimized to minimize evaluating the likelihood in parameter regions that are strongly disfavored.
To that end we start by first considering PDF and FF parameters separately using flat priors, with the resulting samples from each type of hadron structure combined at a later stage to build new prior samples for the final runs.
The workflow is illustrated in \fref{workflow}, where each step is represented as vertical arrows that accumulate additional experimental data from the previous step, with the posterior samples at each step becoming the priors for the subsequent step.
This strategy allows the samples to become more optimized and avoids unnecessary likelihood evaluations in regions of parameters space by disfavoring those regions in earlier stages of the multi-step chain.

\section{Data Sets}
\label{s.datasets}

The data sets used in the present analysis include the primary electromagnetic processes that traditionally have been used in global QCD analyses, namely, inclusive DIS, Drell-Yan lepton-pair production (which constrain PDFs), SIA (which constrains FFs), and SIDIS (which constrains both PDFs and FFs).
The inclusive DIS data are measurements of the $F_2(\xbj,Q^2)$ structure function performed by the BCDMS~\cite{Benvenuti:1989rh, Benvenuti:1989fm} and New Muon Collaborations~\cite{Arneodo:1996qe, Arneodo:1996kd} at CERN, and from experiments at SLAC~\cite{Whitlow:1991uw}, as well as from reduced electron and positron cross sections from the H1 and ZEUS Collaborations~\cite{Abramowicz:2015mha} at DESY.
These include both proton~\cite{Benvenuti:1989rh, Whitlow:1991uw, Arneodo:1996qe} and deuteron~\cite{Benvenuti:1989fm, Whitlow:1991uw, Arneodo:1996kd} targets, and with both neutral and charged current probes~\cite{Abramowicz:2015mha}.
For the kinematics we implement cuts of $W^2 > 10$~GeV$^2$ and $Q^2 > m_c^2$, 
where $W^2 = M^2 + Q^2 (1-\xbj)/\xbj$, in order to select DIS data that can be fitted within leading power factorization.

For Drell-Yan lepton-pair production data we use differential cross section measurements $\diff^2\sigma_{\mbox{\tiny DY}}/\diff{Q}\diff{\xF}$ by the E866/NuSea Collaboration~\cite{Hawker:1998ty, Towell:2001nh, Webb:2003bj} at Fermilab, which include proton scattering from proton and deuteron targets.  
We include data in the range $Q^2 > 36$~GeV$^2$. 
Excluding lower $Q^2$ data is recommended by Ref.~\cite{Alekhin:2006zm}, which demonstrated that inclusion of the lower $Q^2$ data results in deteriorated prediction quality with no reduction in uncertainty when compared with fits to DIS data alone.

\begin{figure}[t!]
\centering
\includegraphics[width=\textwidth]{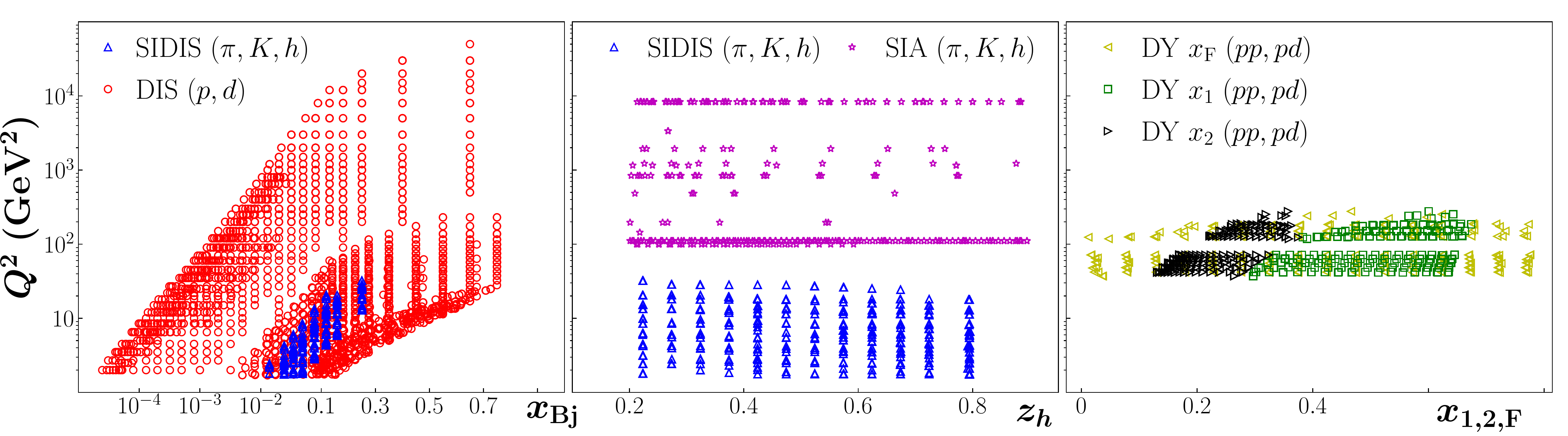}
\caption{Kinematic coverage of data used in this analysis, with $Q^2$ versus the Bjorken scaling variable $\xbj$ for inclusive DIS~\cite{Benvenuti:1989rh, Benvenuti:1989fm, Whitlow:1991uw, Arneodo:1996qe, Arneodo:1996kd, Abramowicz:2015mha} and SIDIS data~\cite{Adolph:2016bga, Adolph:2016bwc} (left panel), fragmentation variable~$z$ for SIDIS and SIA data~\cite{Brandelik:1980iy, Althoff:1982dh, Braunschweig:1988hv, Lu:1986mc, Aihara:1983ic, Aihara:1988su, Cowan:1988gz, Derrick:1985wd, Itoh:1994kb, Abe:2003iy, Buskulic:1994ft, Akers:1994ez, Abreu:1998vq, Albrecht:1989wd, Leitgab:2013qh, Leitgab:2013dva, Lees:2013rqd} (central panel), and momentum fractions $x_1, x_2, \xF$ for Drell-Yan data~\cite{Hawker:1998ty, Towell:2001nh, Webb:2003bj} (right panel).}
\label{f.kinematics}
\end{figure}

All SIA measurements are of the normalized differential cross sections $(\diff\sigma_{\mbox{\tiny SIA}}/\diff{z_h}) / \sigma_{\rm tot}$ for the reaction $e^+ e^- \to (\pi^\pm, K^\pm, h^\pm)\, X$.
The data are from experiments performed by the TASSO~\cite{Brandelik:1980iy, Althoff:1982dh, Braunschweig:1988hv} and ARGUS~\cite{Albrecht:1989wd} Collaborations at DESY, by the TPC~\cite{Lu:1986mc, Aihara:1983ic, Aihara:1988su, Cowan:1988gz}, HRS~\cite{Derrick:1985wd}, SLD~\cite{Abe:2003iy} and BaBar~\cite{Lees:2013rqd} Collaborations at SLAC, by the OPAL~\cite{Akers:1994ez, Abbiendi:1999ry}, ALEPH~\cite{Buskulic:1994ft} and DELPHI~\cite{Abreu:1998vq} Collaborations at CERN, and by the TOPAZ~\cite{Itoh:1994kb} and Belle~\cite{Leitgab:2013qh,Leitgab:2013dva} Collaborations at KEK. 
As shown in \fref{kinematics}, the SIA data cover the large-$Q^2$ region where a leading power description in terms of FFs should be accurate. 
Approximately half of the SIA data points have $Q \approx M_Z$, while the Belle and BaBar $B$ factories have lower $Q \approx 10.5$~GeV.
To ensure applicability of the leading power formalism, the SIA data in our fits are restricted to the range $0.2 < z_h < 0.9$.

Identification of heavy quark flavors for some of the SIA datasets is achieved through measurement of the total energy and momentum in secondary vertices.
The flavor tagged cross sections for a specific flavor $q=c$ or $b$ are particularly sensitive to the $D_q^h$, $D_{\bar q}^h$ and $D_g^h$ fragmentation functions into the observed hadron $h$.
In general, however, care needs to be taken with the precise method for separating primary quark flavors, and there are ongoing discussions regarding the optimal approach to this. 
For more in-depth discussion see, for example, Ref.~\cite{Sato:2016wqj}.

Finally, the critical addition in this work compared with the previous JAM19 analysis~\cite{Sato:2019yez} is the inclusion of unidentified charged hadron data, along with charged pions and kaons, in the SIDIS off deuterium targets from the COMPASS Collaboration~\cite{Adolph:2016bga, Adolph:2016bwc} at CERN.
Since the SIDIS data
    $\diff{\sigma_{\mbox{\tiny SIDIS}}^{h^\pm}}/\diff{Q^2}\diff{\xbj}\diff{z_h}$
are differential in $\xbj$ and $z_h$, they combine information on both PDFs and FFs, which appear in the description of SIA, Drell-Yan, and DIS data.
Furthermore, as illustrated in \fref{kinematics}, the SIDIS data have significant overlap in $\xbj$ and $z_h$ with the $\xbj$ and $\xF$ range of inclusive DIS and Drell-Yan data, respectively, and the $z_h$ range of SIA data, so that the combined analysis constitutes a genuine test of their universality.
For the COMPASS SIDIS data we use the same kinematic cuts on $W^2$ and $Q^2$ as for inclusive DIS, and restrict the fragmentation variable to $0.2 < z_h < 0.8$ in order to exclude data from the target fragmentation region and avoid large-$z$ threshold corrections.

\section{Assessing universality}
\label{s.uni}

Before proceeding to the results of our numerical analysis, we briefly discuss the criteria for universality of the PDFs and FFs and how these are implemented in our analysis.
Extracting parton correlation functions, and using the extractions to test models of parton structure, is a nontrivial inverse problem, the detailed examination of which is beyond the scope of the present paper.
However, a claim that the success of a fit is a measure of the predictive power of the PDFs and FFs requires a number of basic minimal conditions to be met:
\begin{enumerate}
\item The system of unknown correlation functions must be over-constrained, by which we mean that the constraints on unknown correlations imposed by data (or other theoretical constraints such as sum rules) must be greater than the total number of functions involved. 
\item Each unknown correlation function must appear at least twice within the set of factorization formulas relating the correlation functions to physical observables.  
\item There must be reasonable kinematical overlap between the observables so that correlation functions can be compared within similar ranges of parton momentum fractions. 
\end{enumerate}

Using isospin invariance to relate the PDFs in the proton to those in the neutron, we have seven independent PDFs: 
    $f_u$, $f_d$, $f_s$, $f_{\bar{u}}$, $f_{\bar{d}}$, $f_{\bar{s}}$ and $f_g$, 
with PDFs for heavy flavors generated perturbatively.
For the FFs, there are five functions for $\pi^+$ production: 
    $D^{\pi^+}_u$, $D^{\pi^+}_{\bar{u}}$, $D^{\pi^+}_c$, $D^{\pi^+}_b$ and $D^{\pi^+}_g$, 
assuming that for equal $u$ and $d$ quark masses we can equate
    $D^{\pi^+}_{\bar{d}} = D^{\pi^+}_u$.
Charge symmetry allows all the FFs for $\pi^-$ production to be related to those for $\pi^+$ production.
For $K^+$ production, there are six independent FFs:
    $D^{K^+}_u$, $D^{K^+}_{\bar s}$, $D^{K^+}_{\bar{u}}$, $D^{K^+}_c$, $D^{K^+}_b$ and $D^{K^+}_g$, 
where we differentiate between the $u$ and $\bar{s}$ functions.
Again, using charge symmetry the FFs for $K^-$ can be obtained from these six $K^+$ FFs.
Finally, for the unidentified charged hadrons $h^\pm$ or residual FFs, we identify five independent functions:
    $D^{\rm res^+}_u$, $D^{\rm res^+}_{\bar{u}}$, $D^{\rm res^+}_c$, $D^{\rm res^+}_b$ and $D^{\rm res^+}_g$.
This makes then a total of 23 functions to be determined.

The quark and gluon PDFs are constrained by their appearance in several sum rules; in particular, the number sum rules,
\begin{equation}
    \label{e.numberSR}
    \int_0^1\diff{x}\, \big( f_q(x) - f_{\bar q}(x) \big)
    = n_q,
\end{equation}
where $n_u = 2$, $n_d = 1$ and $n_s = 0$, and the momentum sum rule,
\begin{equation}
    \label{e.momentumSR}
    \sum_{i=q,\bar q, g} \int_0^1\diff{x}\, x f_i\parz{x} = 1.
\end{equation}
Note that in \ssref{param} these constraints were specifically used to fix the values of the normalization parameters for several fitted functions.  However, for the purpose of assessing universality, they are simply counted as additional independent equations which include and thus constrain the PDFs.

The data sets discussed in \sref{datasets} also constrain the light quark and gluon PDFs since they appear in expressions for multiple independent observables. 
Counting these and also the four sum rules (\ref{e.numberSR}) and (\ref{e.momentumSR}),
\begin{equation} 
f_i(x) \stackrel{i \neq c,b}{\ \longrightarrow\ }\
\begin{cases}
    6 \qquad \text{DIS}         \\
    2 \qquad \text{Drell-Yan}   \\
    6 \qquad \text{SIDIS}       \\
    4 \qquad \text{sum rules}
\end{cases} 
\end{equation}
there is a total of 18 relations between the light quark PDFs.
The heavy quarks appear in an even greater number of observables.
The light quark fragmentation functions appear in at least one SIA observable and, because of charge conjugation invariance, in 2 SIDIS observables,
\begin{equation}
D_i^{\pi^+}(z) \stackrel{i \neq c,b,g}{\ \longrightarrow\ } 
\begin{cases} 
    1 \qquad \text{SIA} \\
    2 \qquad \text{SIDIS}
\end{cases} 
\end{equation}
and similarly for the kaon and charged hadron fragmentation functions.

For a robust stress-test of universality, there should be reasonable overlap of the ranges in parton momentum fraction for both the PDFs and the FFs.
An indication for how well this is achieved in the current fit can be be gleaned from the kinematical coverage plots shown in \fref{kinematics}.
To lowest order in $\alpha_s$, the kinematical variables $\xbj$, $x_1$, $x_2$ and $z_h$ approximate the parton momentum fractions $x$ and $z$, respectively, while QCD evolution relates all values of $Q^2$.
Figure~\ref{f.kinematics} confirms that PDFs and FFs are both constrained by multiple processes in overlapping regions of momentum fractions.

In summary, our analysis does indeed fulfill the basic criterion for qualifying as a test of universality, and retaining predictive power for the PDFs and FFs more generally.
Note, however, that the momentum sum rule for FFs has not been imposed in the analysis.
Instead, this will be used as a consistency check for the final fit in \sref{analysis}.

\section{Numerical analysis}
\label{s.analysis}

In this section we present the results of our simultaneous Monte Carlo analysis of PDFs and FFs.
We begin with a survey of the fitted cross sections for the various global datasets used in this study, focusing especially on the quality of agreement with the SIDIS and SIA data on $\pi^\pm$ and $K^\pm$, as well as unidentified $h^\pm$ production.
We then present our final fitted PDFs and FFs, and discuss the vital role played by the SIDIS and SIA datasets in particular in constraining the strange quark distribution in the proton.

\subsection{Data and theory agreement}

\begin{figure}[t!]
    \centering
    \vspace*{-0.5cm}
    \includegraphics[width=\textwidth]{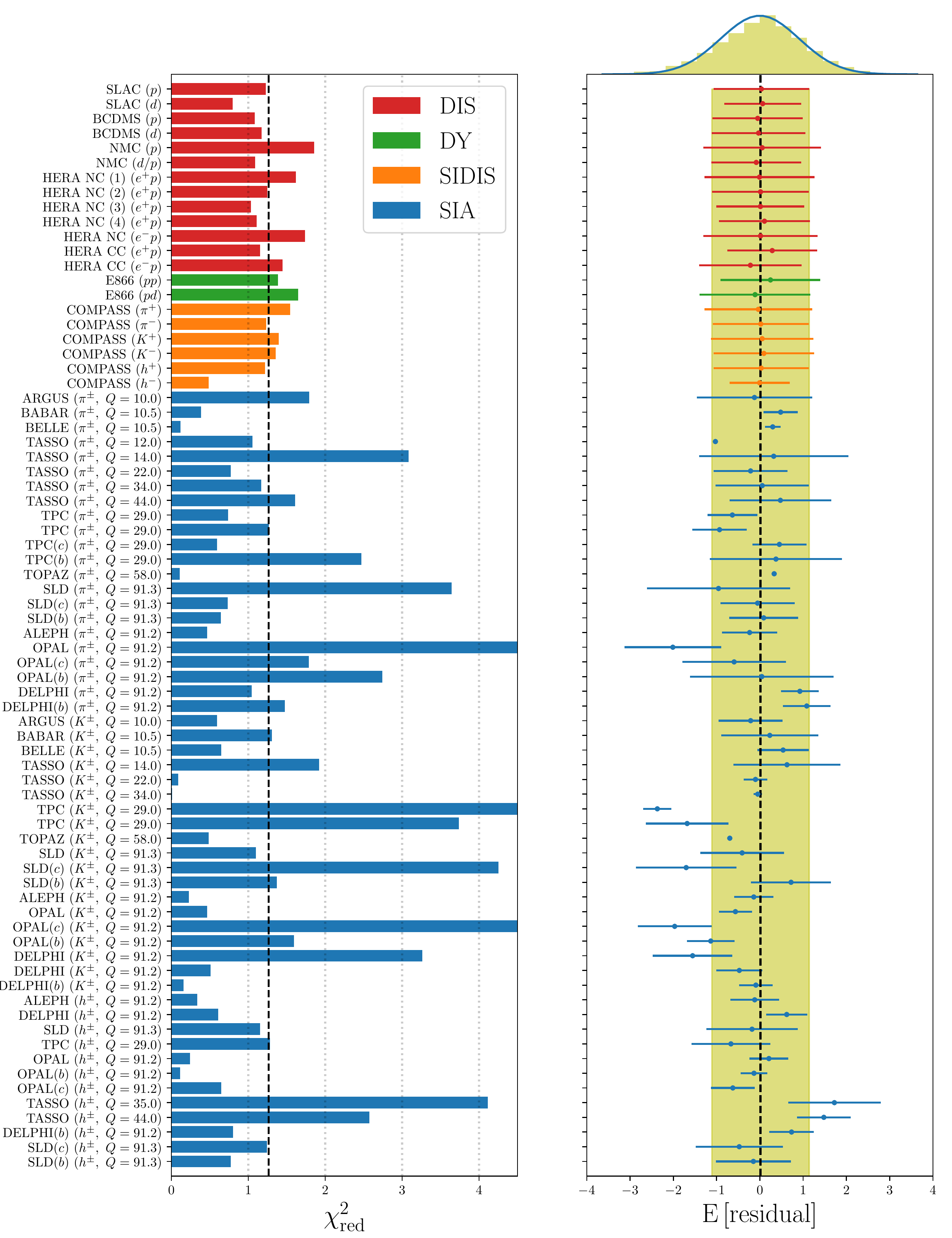}
    \caption{Reduced $\chi^2_{\rm red}$ values for each DIS (red), DY (green), SIDIS (orange) and SIA (blue) experiment considered in this analysis (left column), along with the corresponding  mean and standard deviation of the residuals for each experiment, E\,[residual] (right column).}
    \label{f.chi2}
\end{figure}

To assess the agreement of the fitted results with the various datasets, in \fref{chi2} we show the reduced $\chi^2$ for each individual experiment, which is defined by
\begin{equation}
    \chi^2_{\rm red}
    = \frac{1}{N}\sum_{i,e}
    \frac{1}{\alpha_{i,e}^2}
    \bigg( d_{i,e} - {\rm E}\Big[\sum_k r_e^k \beta^k_{i,e} + T_{i,e}/N_e\Big]
    \bigg)^2.
    \label{e.chi2red}
\end{equation}
Here, the expectation value ${\rm E}[...]$, as defined in \eref{exp2}, represents the mean theory, including optimized multiplicative and additive corrections to match the data, with $N$ the total number of data points.
In \fref{chi2} we show the mean and standard deviation of the Monte Carlo residuals for each experiment $e$, where the residual per data point is defined as
\begin{equation}
    {\rm residual}\,(e,i)
    = 
    \frac{1}{\alpha_{i,e}}
    \bigg( d_{i,e} - {\rm E}\Big[\sum_k r_e^k \beta^k_{i,e} + T_{i,e}/N_e\Big]
    \bigg).
    \label{e.chi2red}
\end{equation}
For the inclusive DIS, Drell-Yan and SIDIS datasets we find excellent overall agreement between data and theory, with $\chi^2_{\rm red}$ values close to 1.
The $\chi^2_{\rm red}$ for the SIA datasets are slightly higher, but nonetheless the overall fit is very good, giving a total reduced $\chi^2_{\rm red}$ = 1.26 for almost 5000 data points.
The values of $\chi^2_{\rm red}$ for each type of dataset and for each specific hadron in the final state are summarized in Table~\ref{t.chi2}, along with the number of data points for each dataset.

\begin{table}[b]
    \centering
    \caption{Reduced $\chi^2_{\rm red}$ values for each type of dataset (DIS, Drell-Yan, SIDIS, SIA) considered in this analysis, together with the number of data points $N_{\rm dat}$ for each dataset.\\}
    \begin{tabular}{ll|c|r} \hline
    \multicolumn{2}{l|}{~~reaction} & ~~~$\chi^2_{\rm red}$~~~ & ~~~$N_{\rm dat}$~~  \\ 
    \hline
    ~~DIS~    &           &      1.29 & 2680~~  \\
    ~~DY~     &           &      1.52&  250~~  \\
    ~~SIDIS~  & $\pi^\pm$~&      1.39 &  498~~  \\
              & $K^\pm$~  &      1.38 &  494~~  \\
              & $h^\pm$~  &      0.85 &  498~~  \\
    ~~SIA     & $\pi^\pm$~&      1.09 &  231~~  \\
              & $K^\pm$~  &      1.37 &  213~~  \\
              & $h^\pm$~  &      1.15 &  120~~  \\
    \hline    
    ~~{\bf total} &       &      1.26 & 4984~~  \\ \hline \\
    \end{tabular}
    \label{t.chi2}
\end{table}

\begin{figure}[t]
    \centering
    \includegraphics[width=0.8\textwidth]{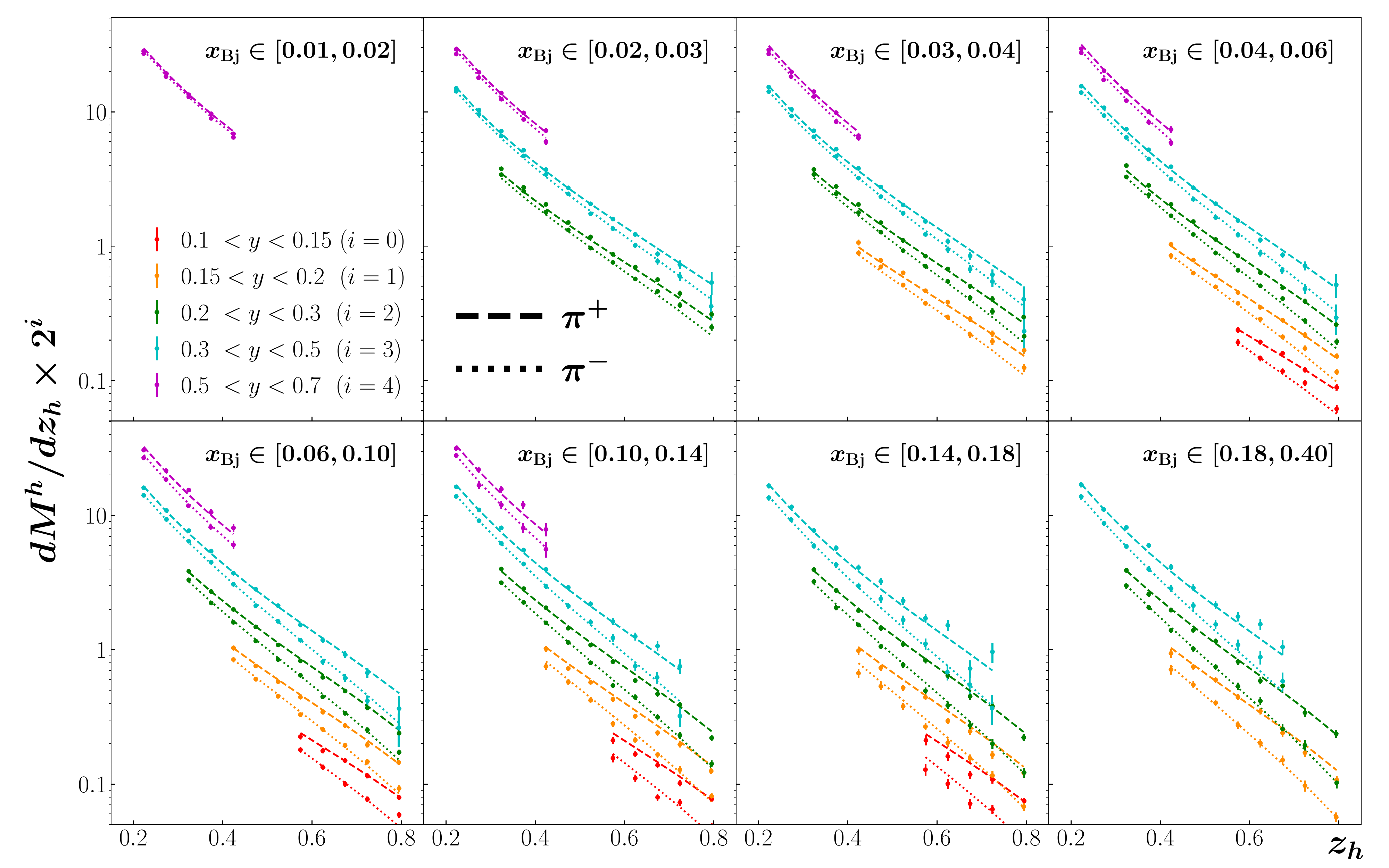}\vspace*{-0.3cm}
    \caption{Comparison of the multiplicities $dM^h/dz_h$ for $h=\pi^+$ (dashed lines) and $\pi^-$ (dotted lines) production with the COMPASS data~\cite{Adolph:2016bga, Adolph:2016bwc} in various bins of $\xbj$ and $y$ (offset by a factor $2^i$).}
    \label{f.sidis-pi}
\end{figure}

\begin{figure}[h]
    \centering
    \includegraphics[width=0.8\textwidth]{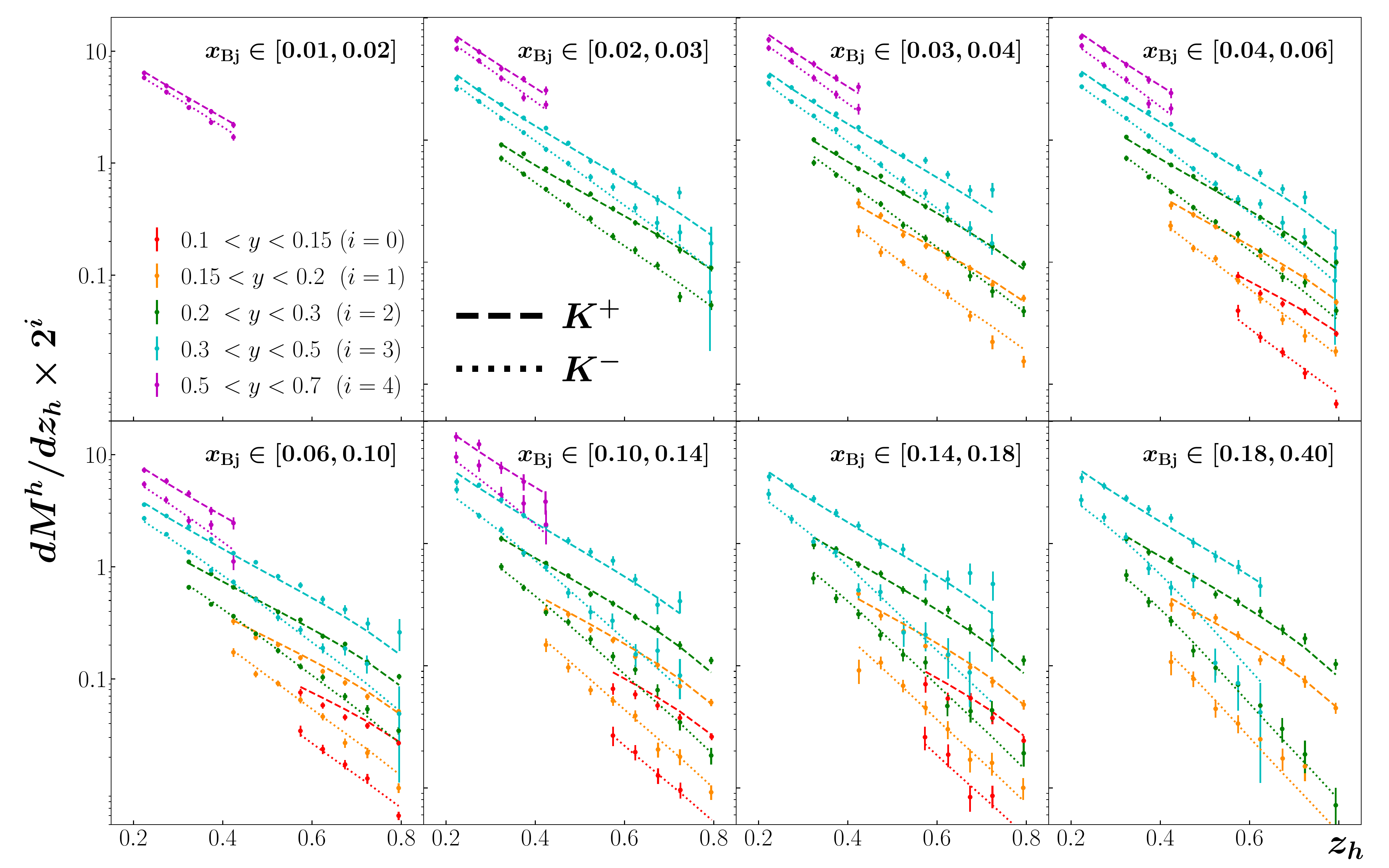}\vspace*{-0.3cm}
    \caption{As in Fig.~\ref{f.sidis-pi}, but for $K^\pm$ COMPASS SIDIS data~\cite{Adolph:2016bga, Adolph:2016bwc}.}
    \label{f.sidis-K}
\end{figure}

\begin{figure}[t]
    \centering
    \includegraphics[width=0.8\textwidth]{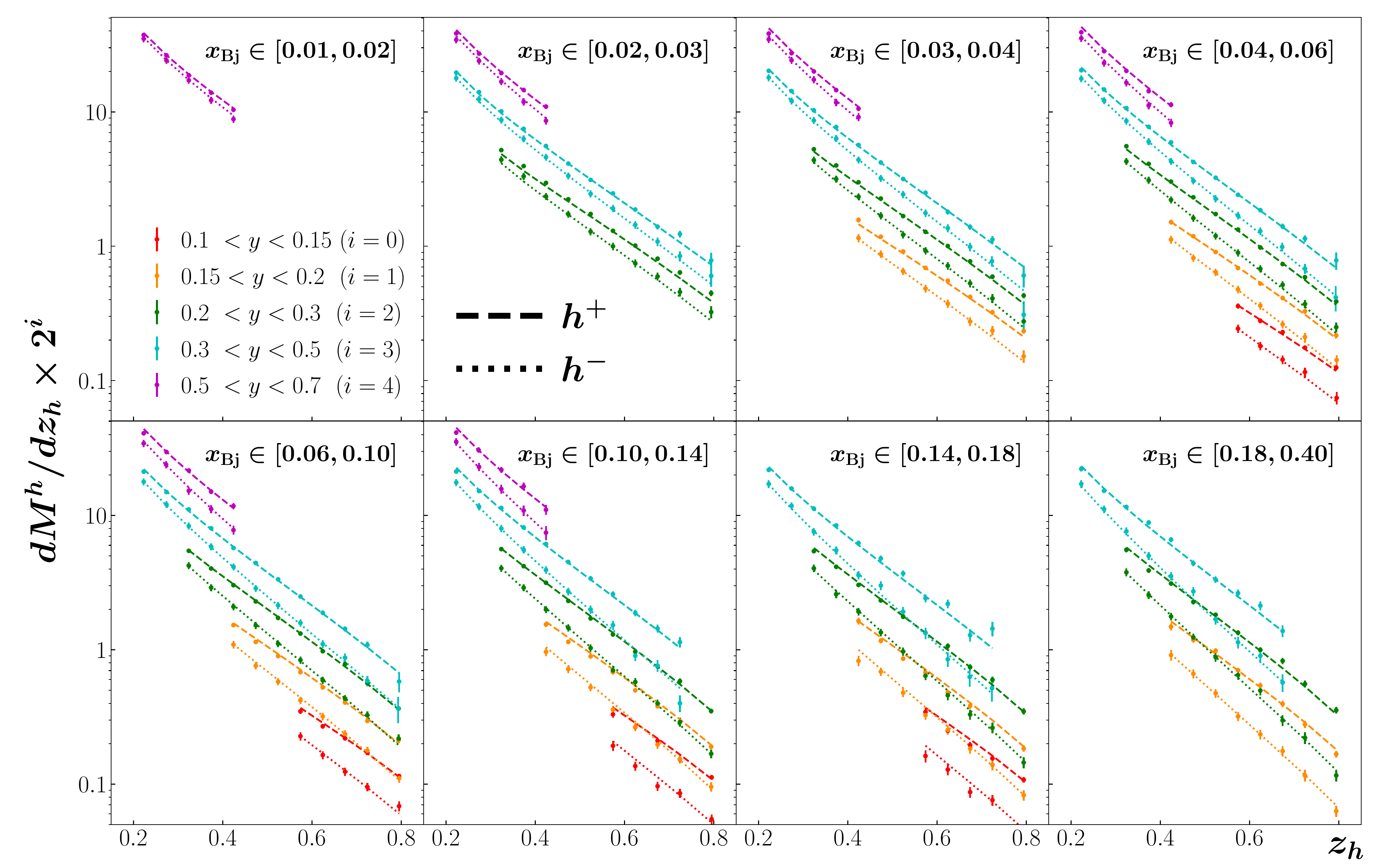}
    \caption{As in Fig.~\ref{f.sidis-pi}, but for unidentified hadron $h^\pm$ COMPASS SIDIS data~\cite{Adolph:2016bga, Adolph:2016bwc}.}
    \label{f.sidis-h}
\end{figure}

The residuals profile for the DIS, Drell-Yan and SIDIS datasets is well centered around zero, with variances $\sim 1$, indicating an average Gaussian behavior of their associated likelihood function.
The variance for the SIDIS $h^-$ data from COMPASS, however, is found to be up to $\approx 50\%$ below unity, suggesting a deviation from a Gaussian likelihood.
This may be due to the fact that these data are dominated by systematic uncertainties, which is also reflected by the relatively small reduced $\chi^2_{\rm red}$ values, especially for the COMPASS $h^-$ data relative to the rest of DIS and SIDIS data sets.

A more detailed comparison with the COMPASS SIDIS is made in Figs.~\ref{f.sidis-pi}, \ref{f.sidis-K} and \ref{f.sidis-h}, where we show the $z_h$ dependence of the $\pi^\pm$, $K^\pm$ and $h^\pm$ multiplicities, respectively, which are defined as ratios of SIDIS to inclusive DIS cross sections at the same $\xbj$ and $Q^2$, 
\begin{equation}
\frac{dM^h}{dz_h}
= \frac{\diff{\sigma_{\mbox{\tiny SIDIS}}^{h}}/\diff{Q^2}\diff{\xbj}\diff{z_h}}
{\diff{\sigma_{\mbox{\tiny DIS}}}/\diff{Q^2}\diff{\xbj}}.
\label{e.multiplicity}
\end{equation}
The agreement between theory and the experimental $z_h$ spectrum is quite remarkable, given that it spans some 2 orders of magnitude, which suggests that at these kinematics a leading power perturbative QCD factorization at next-to-leading order provides sufficient accuracy to describe the data.
Interestingly, the differences between the multiplicities for positively and negatively charged hadron species increase with $\xbj$, especially for kaons, and in the valence region these can differ by an order of magnitude for low values of $Q^2$.
Such differences can enhance our ability to extract flavor dependent effects in nonperturbative PDFs and parton to kaon FFs from the data.
The new data set included for the first time in the present JAM analysis, namely the unidentified charged hadron data shown in \fref{sidis-h}, are also well described by our nonperturbative ansatz for the corresponding FFs.
In contrast to the excellent agreement with the $z_h$ dependence of the data in Figs.~\ref{f.sidis-pi}--\ref{f.sidis-h}, we note that analysis of the same data differential in the hadron transverse momentum using existing PDFs and FFs within TMD factorization results in poor agreement between predictions and data~\cite{Gonzalez-Hernandez:2018ipj, Wang:2019bvb}, indicating that further work is needed to understand the SIDIS transverse momentum spectra.

\begin{figure}[t]
    \centering
    \includegraphics[width=\textwidth]{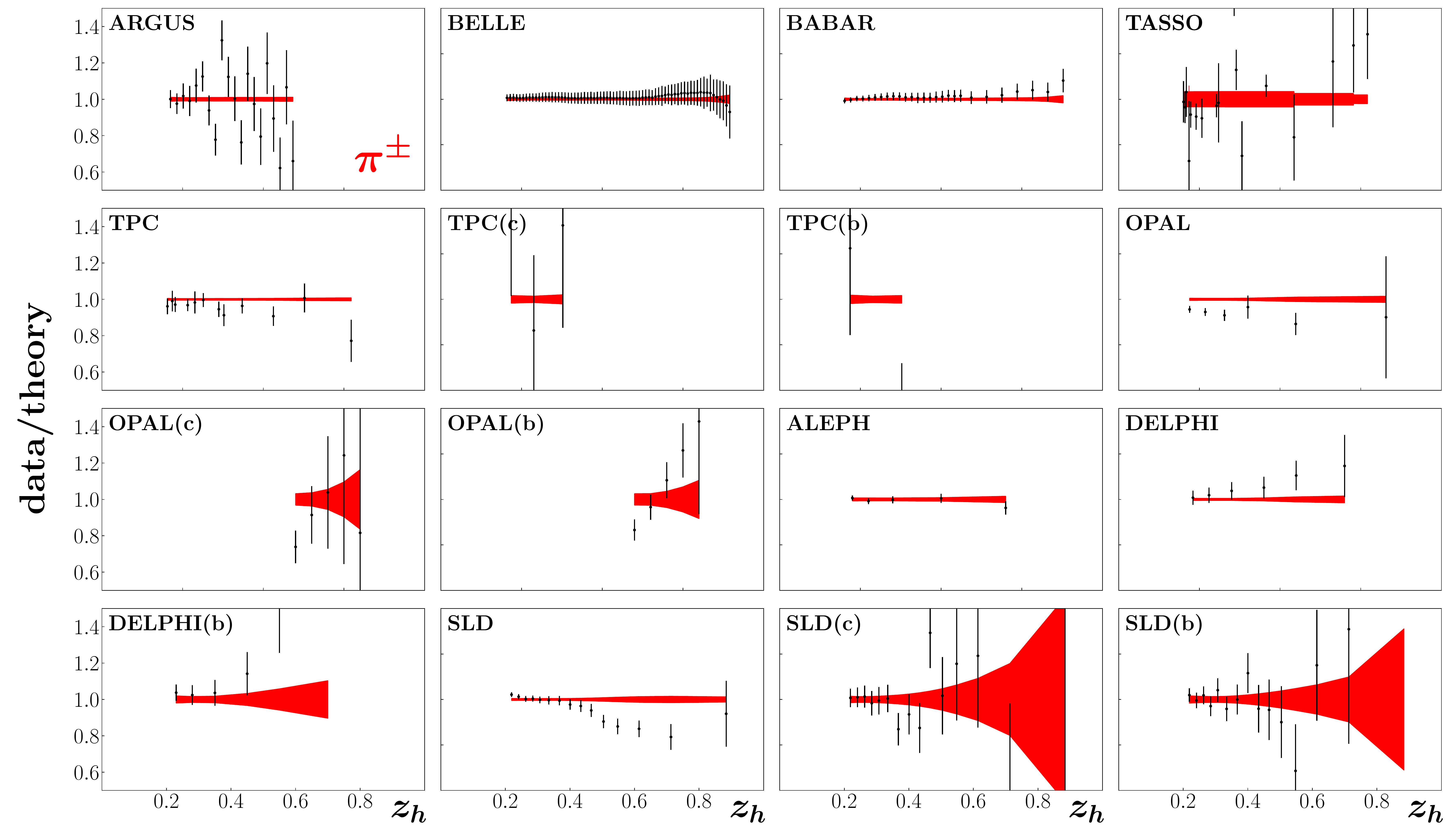}    
    \caption{Data to theory ratios for SIA $\pi^\pm$ production cross sections versus $z_h$, with the bands indicating the uncertainty on the fitted result.}
    \label{f.sia-pi-dot}
\end{figure}
\begin{figure}[h!]
    \centering
    \includegraphics[width=\textwidth]{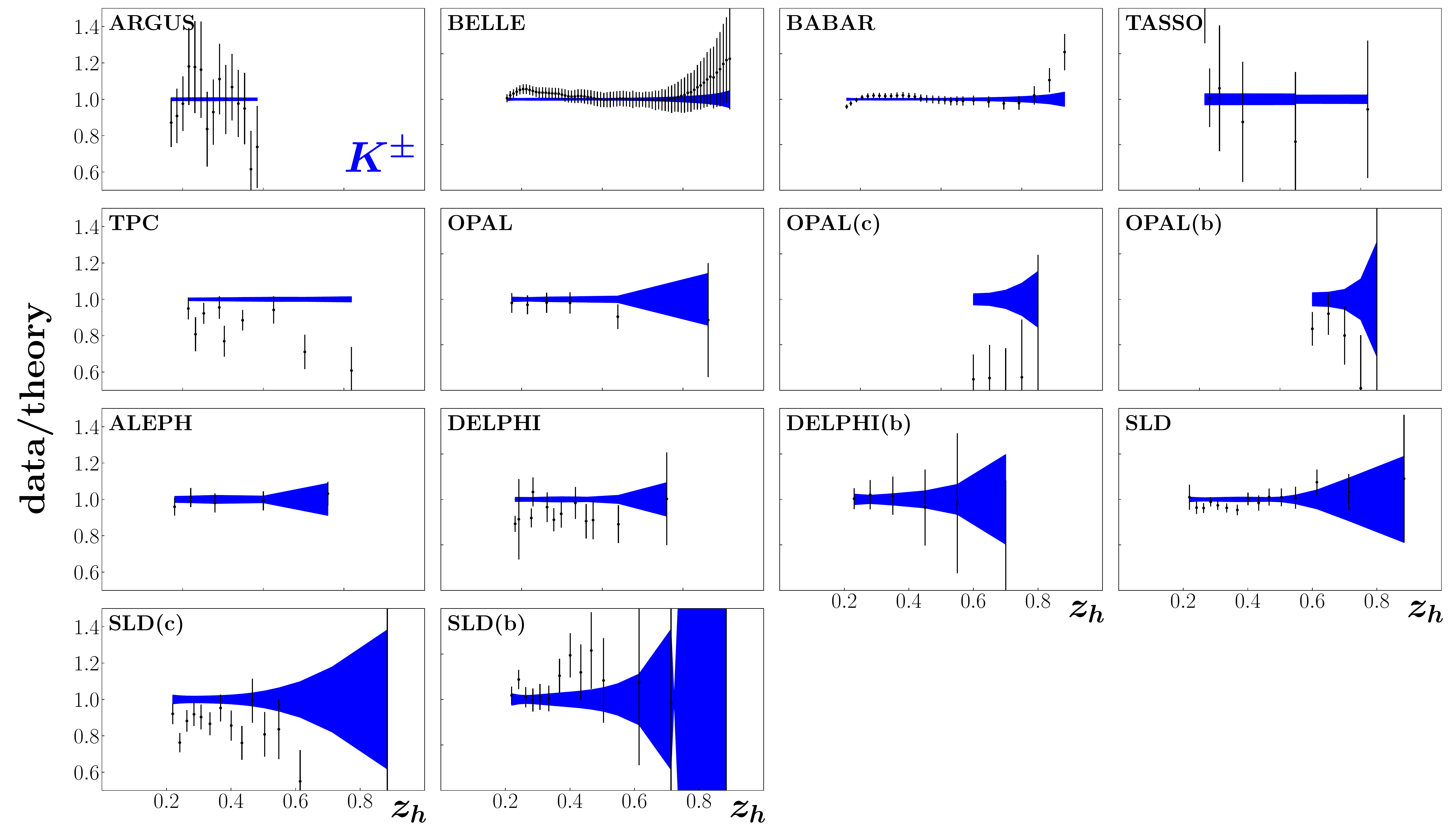}    
    \caption{As in \fref{sia-pi-dot}, but for SIA $K^\pm$ production.}
    \label{f.sia-k-dot}
\end{figure}
\begin{figure}[h!]
    \centering
    \includegraphics[width=\textwidth]{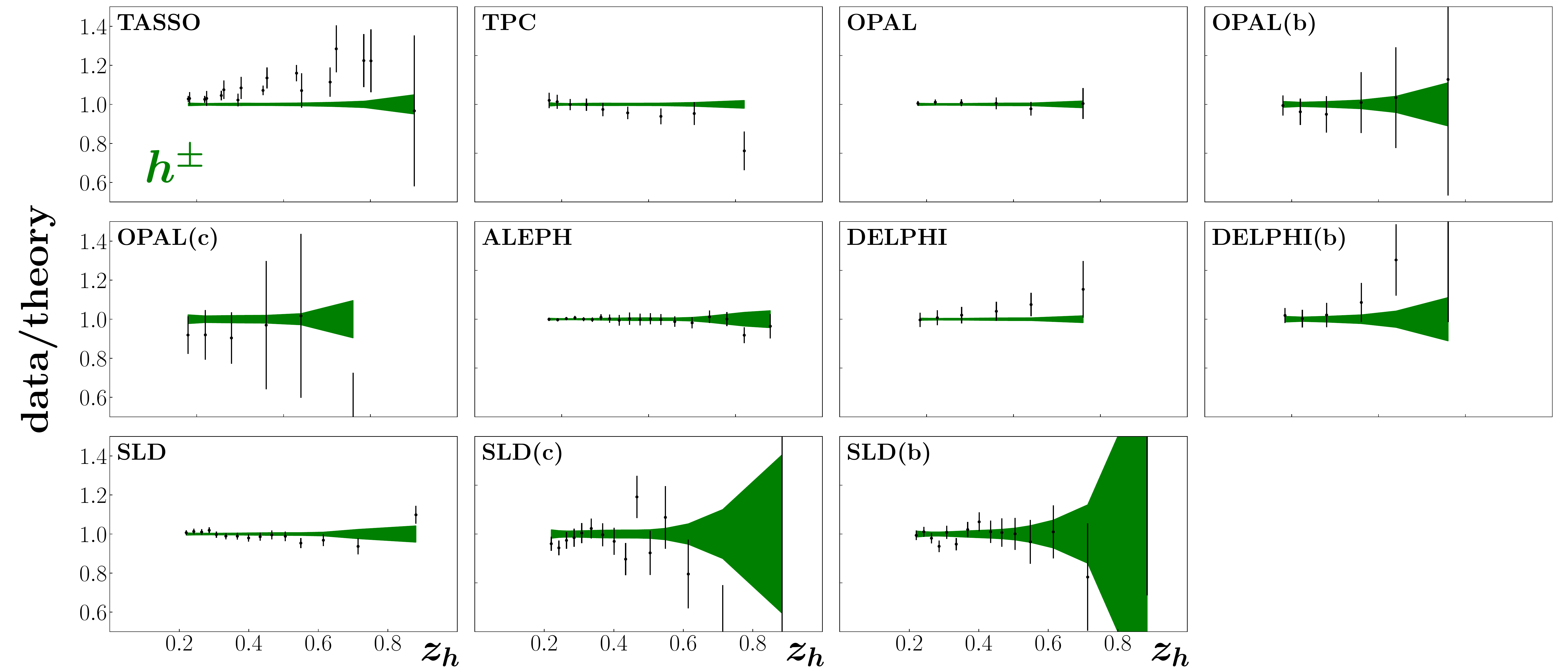}
    \caption{As in \fref{sia-pi-dot}, but for SIA unidentified charged hadron $h^\pm$ production.}
    \label{f.sia-h-dot}
\end{figure}

For the SIA data sets, there is a somewhat wider spread in the data versus theory comparisons, as seen in Figs.~\ref{f.sia-pi-dot}, \ref{f.sia-k-dot} and \ref{f.sia-h-dot} for the $\pi^\pm$, $K^\pm$ and unidentified charged hadron $h^\pm$ final states, respectively.
Generally, the $\pi^\pm$ data have the best agreement among the SIA datasets, with a reduced $\chi^2_{\rm red} = 1.09$, followed by the hadron data with $\chi^2_{\rm red} = 1.15$, and lastly the kaon data, which have an overall reduced $\chi^2_{\rm red} = 1.37$.
For about 3/4 of the $\approx 40$ SIA datasets, we find very good agreement with the global fit, with $\chi^2_{\rm red} \approx 1$ or below.
For the remaining datasets that have larger $\chi^2_{\rm red}$ values, to better understand the reasons for some of the tensions between data and theory we discuss in the following some individual cases ranked by the reduced $\chi^2_{\rm red}$ values.

Starting with the datasets that have the largest $\chi^2_{\rm red}$ values, namely, $\chi^2_{\rm red} \gtrsim 3$, we identify the OPAL ($\pi^\pm$ and $c \to K^\pm$), TPC ($K^\pm$), SLD ($\pi^\pm$ and $c \to K^\pm$), DELPHI ($K^\pm$), and TASSO ($\pi^\pm$ and $h^\pm$ at 35 GeV) datasets.
For the inclusive OPAL ($\pi^\pm$) data, we observe in \fref{sia-pi-dot} that for $z_h < 0.5$ the data are indeed in tension with the corresponding inclusive ALEPH and SLD results, and the overall trend of the data/theory ratio suggests a possible normalization issue with this dataset.  This can also be said for the DELPHI ($K^\pm$) which appears to have some tension with the corresponding inclusive OPAL and ALEPH results.
Similarly, from \fref{sia-k-dot} we find that the TPC ($K^\pm$) spectrum lies below the theory, suggesting again a normalization problem with these data.  
The situation for the TASSO ($\pi^\pm$) data is less clear, as only the $Q=14$~GeV dataset seems to give a bad fit, while data at other energies can be described fairly well.
This again hints at a problem with the overall normalization for this dataset.
The same behavior appears also in the TASSO ($h^\pm$) data in \fref{sia-h-dot}, where both the $Q=35$ and 45~GeV datasets are above the theoretical cross sections.
The case of SLD and OPAL ($c \to K^\pm$) data in \fref{sia-k-dot} shows a clear overestimation of the $z_h$ spectra.
While one can argue that this problem could be a reflection of the need for a more sophisticated heavy quark treatment in our theory, the description of $b$-tagged data from SLD, DELPHI and OPAL is relatively good, so that an explanation in terms of a normalization uncertainty in the SLD and OPAL ($c \to K^\pm$) data may be more relevant.

For SIA datasets that have smaller, but still large, $\chi^2_{\rm red}$ values, $2 \lesssim \chi^2_{\rm red} \lesssim 3$, we identify the $b$-tagged TPC ($b \to \pi^\pm$), OPAL ($b \to \pi^\pm$), and TASSO ($h^\pm$ at 44 GeV).
For the case of the TPC ($b \to \pi^\pm$) data, we see from \fref{sia-pi-dot} that for the largest $z_h$ bin the theory overestimates the data.
On the other hand, good agreement is found for the SLD ($b \to \pi^\pm$) data at the same kinematics.
It is possible that at the smaller $Q$ values of TPC relative to SLD, the range in $z_h$ where leading power factorization is applicable is narrower, in particular for the $b$-tagged data. 
The $z_h$ dependence of the OPAL ($b \to \pi^\pm$) data appear to be clearly different from the theory, even within the large uncertainties.
We note here that the OPAL data are presented as truncated moments as a function of the lower limit of the integration, $z_h^{\rm min}$, and the inclusion of the very high $z_h$ bins may be problematic for the validity of factorization theorems at $z_h \to 1$.
Lastly, as with TASSO ($h^\pm$ at 35 GeV), the somewhat large $\chi^2_{\rm red}$ values for the 44 GeV data is likely attributable to a problem with overall normalization.

For datasets that have $\chi^2_{\rm red} \lesssim 2$, we consider the agreement to be generally acceptable.
Indeed, the vast majority of datasets in this category have $\chi^2_{\rm red} \approx 1$ or below.
These include all of the recent high-statistics $B$-factory data from BaBar ($\pi^\pm$, $K^\pm$) and Belle ($\pi^\pm$, $K^\pm$), most of the TASSO ($\pi^\pm$, $K^\pm$), TPC ($\pi^\pm$, $c \to \pi^\pm$) and SLD ($h^\pm$, $b \to \pi^\pm$, $b \to h^\pm$) datasets, all of the ALEPH ($\pi^\pm$, $K^\pm$, $h^\pm$) and most of the DELPHI ($\pi^\pm$, $K^\pm$, $b \to K^\pm$, $h^\pm$, $b \to h^\pm$) data, along with the older ARGUS ($K^\pm$), TOPAZ ($\pi^\pm$, $K^\pm$) and OPAL ($K^\pm$, $h^\pm$, $c \to h^\pm$, $b \to h^\pm$) data.
Slightly higher, but still reasonable, $\chi^2_{\rm red}$ values are obtained for the ARGUS ($\pi^\pm$), TPC ($h^\pm$), DELPHI ($b \to \pi^\pm$), and SLD ($K^\pm$, $c \to \pi^\pm$, $c \to h^\pm$) datasets.

Finally, we note that most of the large $\chi^2_{\rm red}$ values found in this analysis were absent in the previous JAM Monte Carlo analysis of fragmentation functions~\cite{Sato:2016wqj}.
The main reason is the restriction of the SIA datasets here to the range $0.2 < z_h < 0.8$, chosen to coincide with the range over which the SIDIS data in this work are able to be described within collinear factorization.
For the LEP data in particular there are many data points at $z_h < 0.2$ which can be well fitted within the current framework, and which would reduce the overall $\chi^2_{\rm red}$.
A~careful point by point comparison of the individual $\chi^2_{\rm red}$ values for the various datasets indeed confirms that similar discrepancies also occurred in Ref.~\cite{Sato:2016wqj}.
However, for consistency in our joint analysis of PDFs and FFs, we restrict the kinematic range to the region where both SIA and SIDIS can be simultaneously described.
The same choice for the $z_h$ range was made in the recent JAM19 analysis, which required SIDIS data to be restricted to $z_h \gtrsim 0.2$ to ensure separation of the target and current fragmentation regions.

\subsection{Parton distributions and fragmentation functions}
\label{ss.pdfsff}

The proton PDFs from our simultaneous fit are displayed in \fref{pdfs} at a scale $\mu^2=10$~GeV$^2$, where we focus on the kinematic region of parton momentum fractions $x \gtrsim 0.01$ that is constrained by the SIDIS data.
For comparison, we contrast our results with other next-to-leading order PDF parametrizations, namely, from the CJ15~\cite{Accardi:2016qay} and NNPDF3.1~\cite{Ball:2017nwa} global analyses.
Compared with the other fits, our valence $u$ and $d$ quark distributions have slightly larger magnitude in the intermediate-$x$ region, $x \sim 0.1$, with a compensating stronger suppression at small $x$ needed to ensure that the valence number sum rules are respected.
The ratio $d/u$ is quite compatible with the results from the other groups, on the other hand, but has a significantly larger uncertainty at large $x$ compared with the CJ15 result, reflecting the Monte Carlo nature of our analysis.

\begin{figure}
    \centering
    \includegraphics[width=\textwidth]{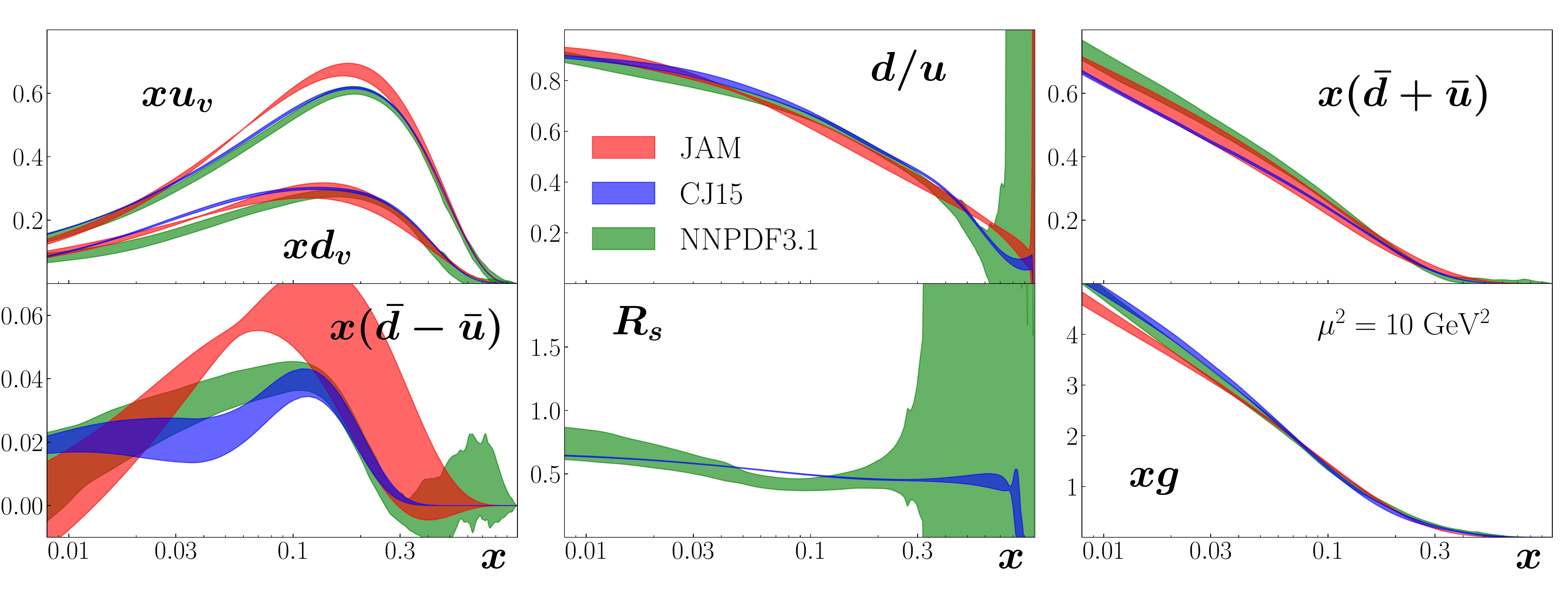}
    \caption{Proton PDFs from the present JAM20-SIDIS analysis (red bands) versus $x$ at a scale $\mu=10$~GeV$^2$, compared with the CJ15~\cite{Accardi:2016qay} (blue bands) and NNPDF3.1~\cite{Ball:2017nwa} (green bands) parametrizations.  The bands shown represent the mean $\pm 1\sigma$.}
    \label{f.pdfs}
\end{figure}

The intermediate-$x$ enhancement in the valence PDFs in our fit is correlated with the slightly smaller $\bar d+\bar u$ light antiquark sea compared with the CJ15 and NNPDF3.1 parametrizations.
This in turn is correlated with the behavior of the strange quark sea, as seen in the ratio
\begin{equation}
R_s = \frac{s+\bar s}{\bar d+\bar u}
\end{equation}
of the strange to nonstrange sea quark PDFs.
In \fref{pdfs} this ratio is generally larger in our analysis than for the other parametrizations, with a somewhat bigger uncertainty.
This is understood from the fact that in the CJ15 fit $R_s$ is fixed to be 0.4 at the input scale, with deviations from the constant value arising only from DGLAP evolution.
For the NNPDF3.1 fit the uncertainties are smaller because of their inclusion of the neutrino DIS data, which we do not include in our analysis because of unknown nuclear corrections in neutrino scattering~\cite{Accardi:2009qv, Kovarik:2010uv, Kalantarians:2017mkj}. 
Our light antiquark asymmetry $\bar d-\bar u$ is also compatible with the other groups, but again with a larger uncertainty, which may be related to the absence of collider $W$ and lepton asymmetry data in our fit.
Finally, for the gluon distribution, the magnitude and uncertainties are very similar across all the analyses, even though our fit does not include jet production data from hadron colliders.
This reflects the fact that the HERA DIS data, which are included here, provide strong constraints on the shape of the gluon PDF via scaling violations.

\begin{figure}
    \centering
    \includegraphics[width=\textwidth]{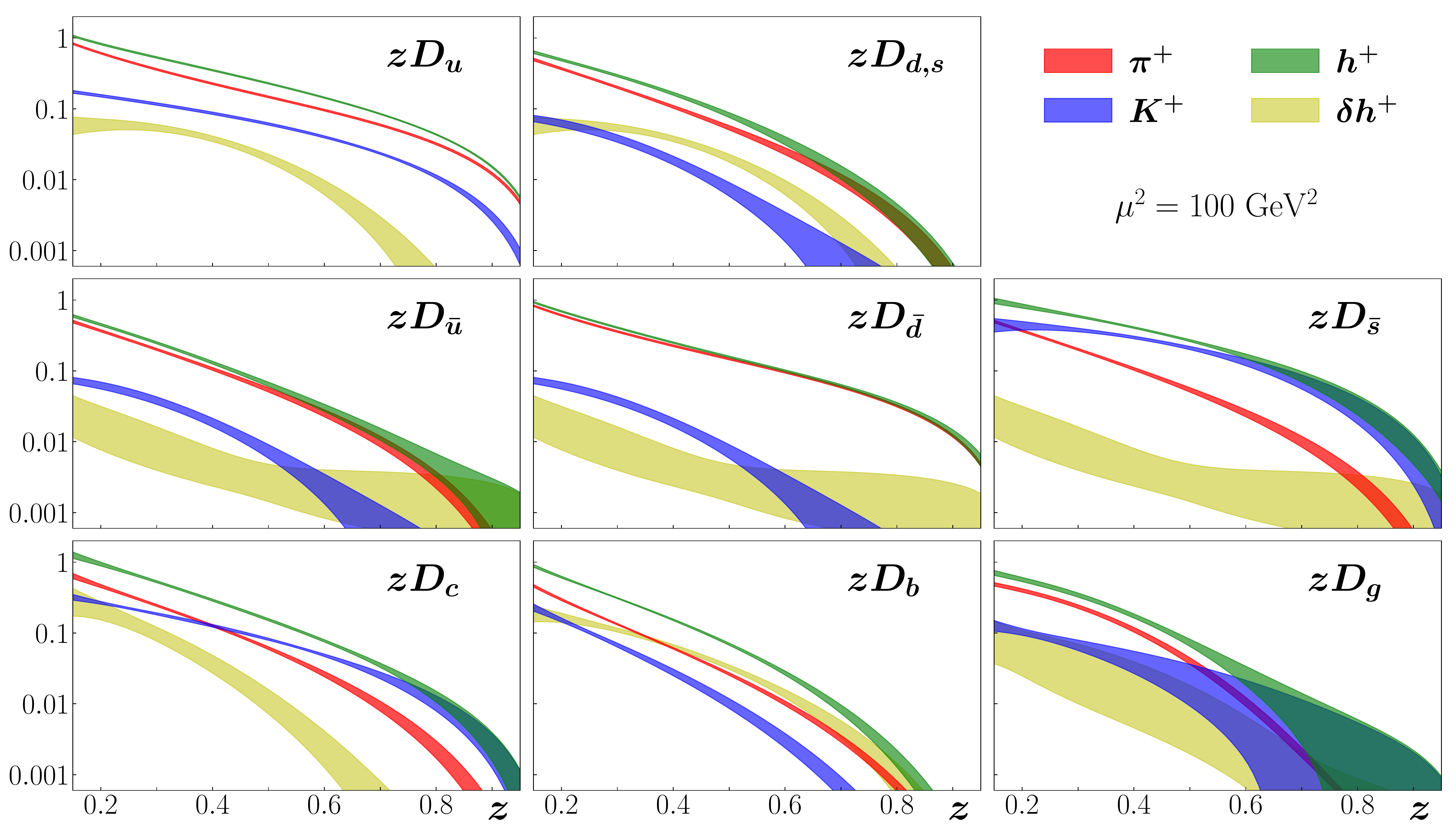}
    \caption{Parton to hadron FFs versus $z$ at $\mu^2=100$~GeV$^2$ from the JAM20-SIDIS analysis for various parton flavors fragmenting to $\pi^+$ (red bands), $K^+$ (blue bands), unidentified hadrons $h^+$ (green bands), and residual hadrons $\delta h^+$ (yellow bands), defined as the difference between $h^+$ and the sum of $\pi^+$ and $K^+$.  The bands shown are mean $\pm 1\sigma$.}
    \label{f.ffs}
\end{figure}

For the parton to hadron FFs, we show in \fref{ffs} the $z$ dependence of the FFs at a scale $\mu^2=100$~GeV$^2$ for the positively charged $\pi^+$, $K^+$ and unidentified hadrons $h^+$, as well as for the residual hadrons $\delta h^+$, defined as the difference between $h^+$ and the sum of $\pi^+$ and $K^+$ (so that the total is given by $h^+ = \pi^+ + K^+ + \delta h^+$).
For most of the flavors we find that the quark $\to$ $\pi^+$ fragmentation dominates, as expected from the pion being the lightest hadron in the QCD spectrum. 
Exceptions to this are for $\bar{s} \to K^+$ and $c \to K^+$ at intermediate $z$ values, and for $b$ quark fragmentation into residual hadrons $\delta h^+$.

For gluon fragmentation, pion production dominates for $z$ up to $\sim 0.5-0.6$, above which kaon fragmentation becomes as sizeable as the pion. 
This is consistent with the findings of previous FF analyses~\cite{Hirai:2007cx, Sato:2016wqj}, which observed that the production of heavier particles such as kaons requires larger momentum fractions from the fragmenting gluon compared to the production of lighter particles.

The production of hadrons heavier than kaons, as indicated in \fref{ffs} by the residual hadrons $\delta h^+$, can be sizable and comparable to that of kaons, especially for the $d$ and $s$ quarks and at large values of $z$.
The relatively large $d \to \delta h^+$ FF can be understood in terms of the fragmentation into protons.
Note that we have imposed flavor symmetry for the residual hadron fragmentation, so that $D_d^{\delta h^+} = D_s^{\delta h^+}$.
In principle, the presence of hyperons such as $\Sigma^+$ should brake this relation, but we leave analysis of such effects for future work.
As the case for the $g \to K^+$, the fragmentation of gluons into heavier particles peaks at large $z$, where larger momentum fractions from the fragmenting gluons are need for the production of heavier particles.

For production of hadrons initiated by heavy quarks, we find similar fragmentation of charm quarks into pions and kaons, but a rather different pattern for the fragmentation of bottom quarks.
Some of this difference can be explained by the flavor-changing properties of $u$-type quarks decaying into $d$-type quarks.
While the charm quark can decay into strange quarks and hence enhance $K^+$ production, the same does not occur for bottom quarks, which suppresses kaon production relative to pion production due to the mass difference. 
Interestingly, the production of other species of charged hadrons is much larger for $b$ quarks than for $c$ quarks, which may be understood from the greater phase space available for $b$ quarks to decay into heavier hadrons to which charm quarks cannot transition.

\begin{figure}
    \centering
    \includegraphics[width=\textwidth]{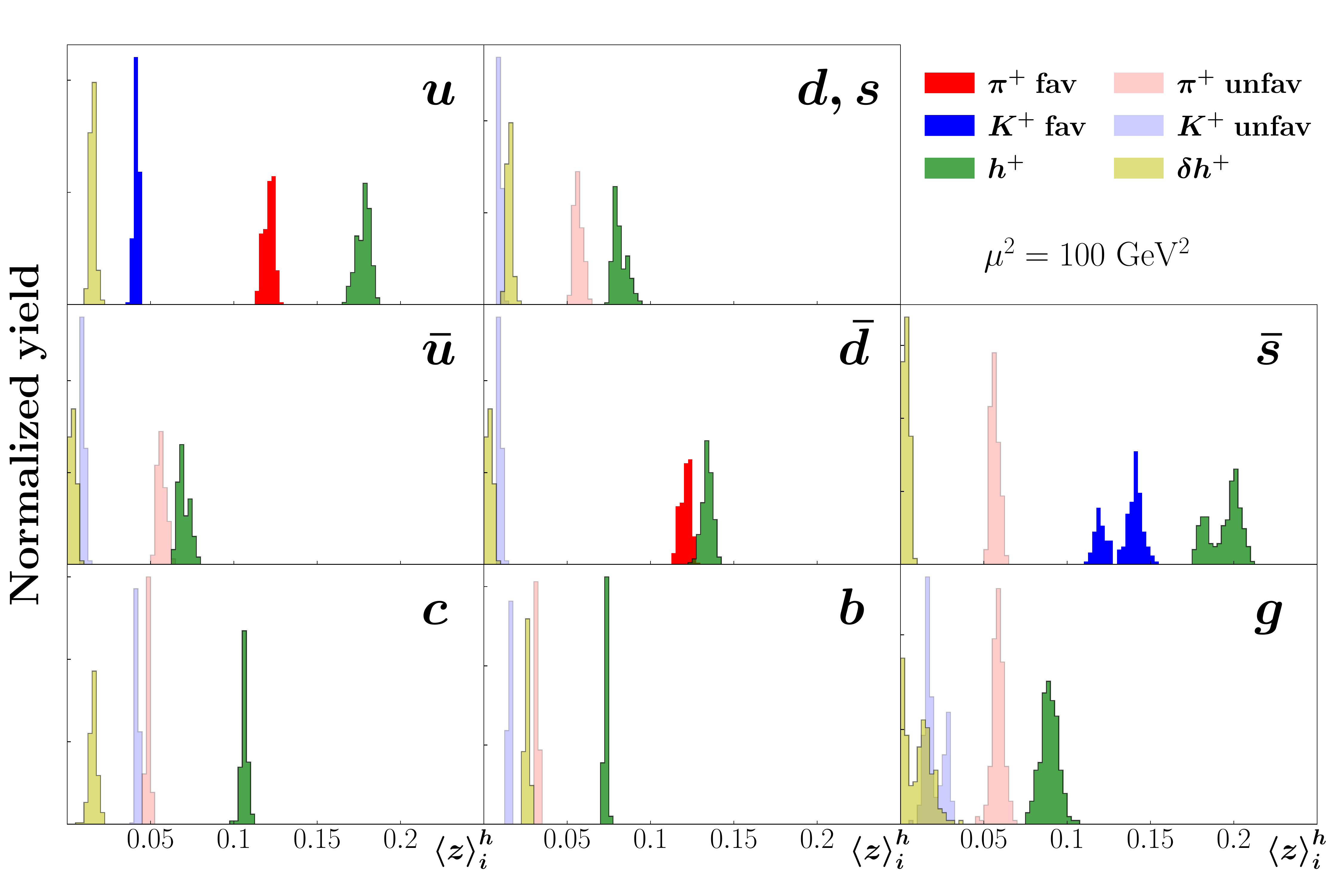}    
    \caption{Normalized yield of truncated moments 
    $\langle z \rangle_i^h$ of the $i \to h$ FFs $z D_i^h$, for the favored $\pi^+$ (red) and favored $K^+$ (blue), unfavored $\pi^+$ (light red) and unfavored $K^+$ (light blue), the total hadron $h^+$ (green) and residual hadron $\delta h^+$ (yellow) FFs, at a scale of $\mu^2=100$~GeV$^2$.}
    \label{f.sumrule}
\end{figure}

In \fref{sumrule} we present truncated moments
\begin{equation}
\label{e.ztrunc}
\langle z \rangle_i^h
= \int_{z_{\rm min}}^1\, dz\, z D_i^h(z),
\end{equation}
for each flavor $i$ and final state hadron $h$, where we take the lower limits on the $z$ integration $z_{\rm min} = 0.2$ to restrict the moment to the region of SIDIS kinematics.
The truncated moment indicate how energetic is the production different a hadron species $h$ relative to the parent parton $i$.
In general, we find that the production of hadron species heavier  than pions and kaons is typically produced with lower energies, which is consistent with the physical picture whereby more energy is required to produce heavier hadrons than lighter hadrons.

As expected, the favored fragmentation of $\bar{d}$ quarks is predominantly into highly energetic pions, while for the antistrange $\bar s$ the production rate of energetic kaons is slightly higher than that of pions.
The unfavored fragmentation of $d, s$ and $\bar u$ quarks follows a similar pattern, with the lightest (pion) state produced at the highest energies followed by kaons and other heavier charged hadrons.
An exception to this behavior is for charm and bottom quark fragmentation: for $c$ quarks kaons are produced with energies comparable to those of pions, while for $b$ quarks kaon production is suppressed with heavier mass hadrons produced at similar energies as pions.

Interestingly, the production of hadrons from gluons follows the same pattern as for $u$-quark fragmentation.
While the latter can be explained in terms of mass differences between the produced hadron species, the fact that $u$ quarks and gluons give a similar average energy profile across hadron species is intriguing. 
On perturbative grounds one can argue that gluon fragmentation is enhanced because of the $C_A=3$ factor in the the gluon splitting function, $P_{gg}$, relative to quark splitting functions, $P_{qq}$ and $P_{gq}$, which are proportional to $C_F=4/3$.
The absence of direct constraints on the gluon FF beyond scaling violations, however, anything drawing more than speculative conclusions at present.

\begin{figure}
    \centering
    \includegraphics[width=0.9\textwidth]{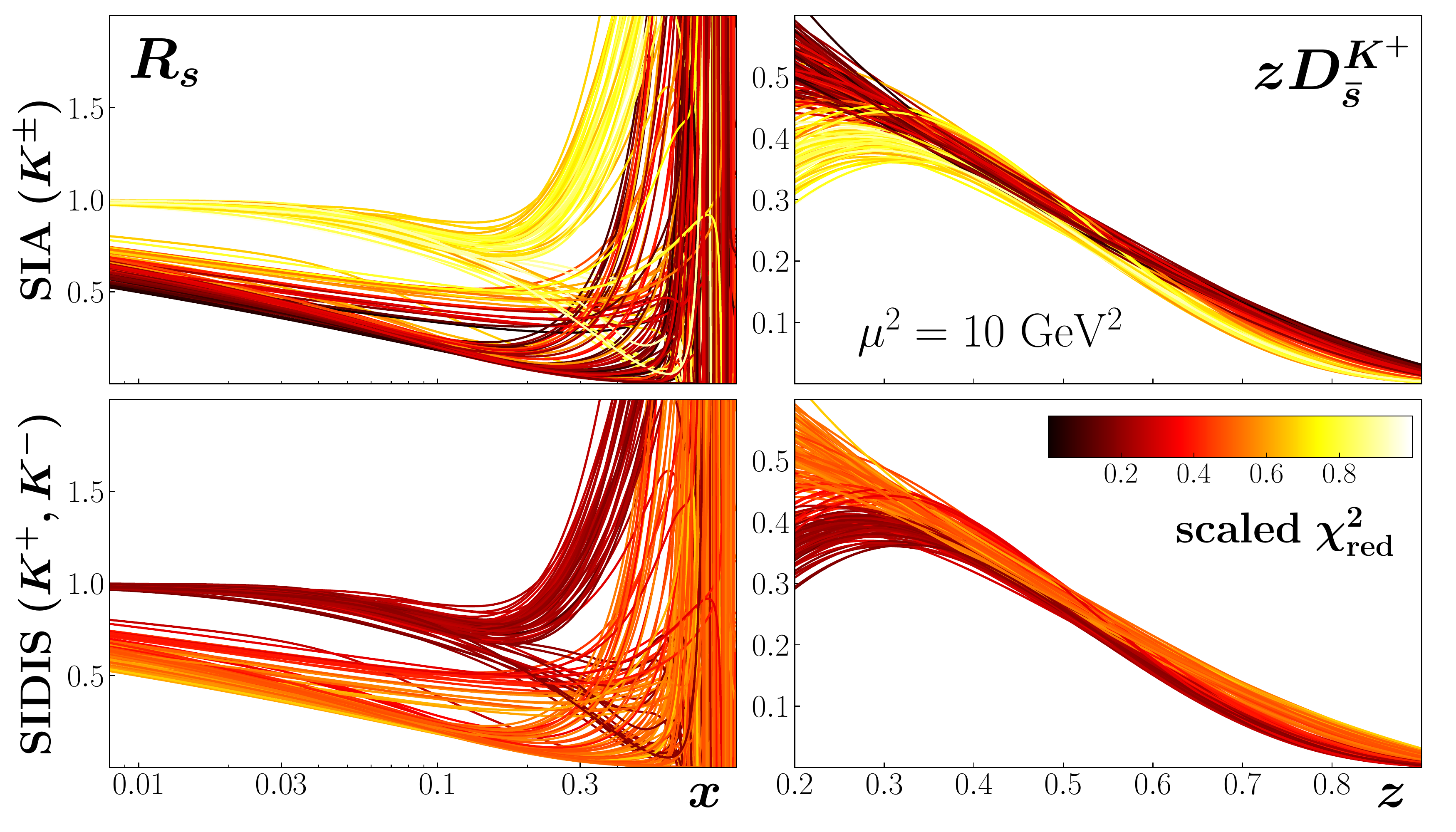}    
    \caption{Monte Carlo samples for the $R_s$ ratio (left) and $zD_{\bar s}^{K^+}$ FF (right) at $\mu^2=10$~GeV$^2$, color coded according to the scaled $\chi^2_{\rm red}$ for the SIA ($K^\pm$) (top row) and SIDIS ($K^+, K^-$) (bottom row) datasets.}
    \label{f.PDFstrange}
\end{figure}

We conclude the discussion of our numerical results by focusing on the correlation between the strange to nonstrange PDF ratio $R_s$ and the strange to kaon fragmentation function $D_{\bar s}^{K^+}$.
In \fref{PDFstrange} we show $R_s$ and the $\bar{s}\to K^+$ FF, with individual Monte Carlo samples color coded by the scaled $\chi^2_{\rm red}$ intensity (with darker replicas indicating higher likelihoods) computed for the specific cases of SIA ($K^\pm$) and SIDIS ($K^+, K^-$) datasets.
The SIA datasets have a clear preference for a smaller $R_s$ and enhanced $D_{\bar{s}}^{K^+}$, as was found in the previous JAM19 analysis~\cite{Sato:2019yez}.
Interestingly, the SIDIS ($K^+, K^-$) data, which have smaller $\chi^2_{\rm red}$, have a slight tendency to favor solutions with a larger $R_s$ and smaller $D_{\bar{s}}^{K^+}$, however, this preference is much weaker than the preference of the SIA data for smaller $R_s$ values.

We also note that in the current analysis we have extended the flexibility of the PDF and FF parametrizations, which allowed us to obtain a more uniform Monte Carlo distribution of $R_s$ compared JAM19, where a more restricted parametrization gave rise to multiple solutions.
Our new analysis confirms that the most probable solutions found in JAM19 did not result from parametrization bias, and corroborates the need for a suppressed strange quark PDF in the proton in order to simultaneously describe both the SIA and SIDIS datasets within leading power QCD factorization.

\section{Conclusion}
\label{s.conclusion}

In this paper we have presented the results of a simultaneous Monte Carlo analysis of PDFs and FFs constrained by a diverse array of data from inclusive and semi-inclusive DIS, Drell-Yan lepton-pair production, and SIA in $e^+ e^-$ collisions.
The analysis extends the previous JAM19~\cite{Sato:2019yez} simultaneous fit by including in addition unidentified charged hadrons in the final states of SIDIS and SIA, and increasing the flexibility of the PDF and FF parametrizations.

The analysis --- referred to as ``JAM20-SIDIS'' --- represents the most comprehensive determination of parton to hadron ($\pi^\pm$, $K^\pm$, $h^\pm$) FFs fitted concurrently with spin-averaged parton distributions, broadening the test of universality of parton correlation functions to more observables.
The more thorough exploration of the parameter space and reduced $\chi^2_{\rm red}$ values for each of the $\approx 70$ datasets fitted in this study confirmed the previous finding~\cite{Sato:2019yez} that the combination of SIA and SIDIS datasets have a strong preference for a smaller strange to nonstrange PDF ratio, $R_s$, correlated with an enhanced $D_{\bar{s}}^{K^+}$ FF.
As further tests of this scenario, we plan in future to extend the experimental datasets to include weak-boson and jet production in hadronic collisions, from both Tevatron and LHC data, as well as to relax the $W^2$ cuts for inclusive DIS to incorporate more fixed-target DIS data at high $\xbj$ values~\cite{Cocuzza21}.

An important application of the current results will be in benchmark calculations of transverse momentum dependent cross sections, and in particular for the small transverse momentum region where the transition from collinear factorization to TMD factorization is expected to set in.
One motivation for the present project was to assess the possible role of limitations in collinear PDF and FF fits in explaining discrepancies between theory and data in the range of intermediate and large transverse momentum across a number of transversely differential processes~\cite{Gonzalez-Hernandez:2018ipj, Wang:2019bvb, Bacchetta:2019tcu, Moffat:2019pci}.
For this, a truly simultaneous analysis of parton distribution and fragmentation functions across the standard set of electromagnetic processes, integrated over all transverse momentum, is necessary.
The general success of the collinear fits for transverse momentum integrated cross sections that we have examined here, and their evident predictive power, suggests that factors unique to the transverse momentum differential treatment are responsible for the tension with data.
We plan to address this also in future work. \\

\begin{acknowledgments}
This work was supported by the US Department of Energy contract DE-AC05-06OR23177, under which Jefferson Science Associates, LLC operates Jefferson Lab.  
The work of N.S. was supported by the DOE, Office of Science, Office of Nuclear Physics in the Early Career Program.
E.M. and T.R. were supported by the U.S. Department of Energy, Office of Science, Office of Nuclear Physics, under Award Number DE-SC0018106.
\end{acknowledgments}

\appendix
\section{Semi-inclusive DIS}
\label{a.sidisrev}

In this appendix we summarize the basic cross section and structure function formulas relevant for our analysis of the semi-inclusive leptoproduction of a hadron $h$ (with four-momentum $p_h$) in the deep-inelastic scattering of a lepton $\ell$ (momentum $l$) from a nucleon $N$ (momentum $p$) via the exchange of a virtual photon (momentum $q$),
\begin{equation}
\ell + N \rightarrow \ell' + h + X\, , 
\end{equation}
where the final state hadron is integrated over all transverse momentum.
The formal setup follows standard methods described in Refs.~\cite{Altarelli:1979kv, Nason:1993xx, Furmanski:1981cw, Graudenz:1994dq}, for example, and we closely follow the specific techniques in Ref.~\cite{deFlorian:1997zj} utilizing the double Mellin moment method from Ref.~\cite{Stratmann:2001pb}.

The spin-averaged cross section is parametrized in terms of the semi-inclusive structure functions $F_1^h$ and $F_L^h$,
\begin{equation}
\frac{\diff{\sigma}{}}{\diff{\xbj} \diff{y}\, \diff{z_h}}
= \frac{4 \pi \alpha^2}{Q^2} 
  \bigg[ \frac{1 + (1-y)^2}{y} F_1^h(\xbj,z_h,Q^2) 
        + \frac{(1-y)}{\xbj\, y}\, F_L^h(\xbj,z_h,Q^2)
  \bigg],
\end{equation}
which are functions of the Bjorken scaling $\xbj$, the hadron fragmentation scaling variable $z_h$, and the four-momentum transfer squared $Q^2$, defined in the standard way as
\begin{equation}
\xbj = \frac{Q^2}{2 p \cdot q}\,, \qquad 
z_h = \frac{p_h \cdot p}{q \cdot p}\,, \qquad 
Q^2 = -q^2,
\end{equation}
with $y = q \cdot p/l \cdot p$ the inelasticity.
Our conventions for the semi-inclusive structure function definitions are directly related to the conventions for the inclusive DIS structure functions, with 
    $F_T = 2 x F_1$
and $F_L = F_2 - 2 x F_1$
the transverse and longitudinal structure functions, respectively.
(Note that other conventions use instead 
    $F_1 \to 2 F_1$ and 
    $F_L \to F_L/x$~\cite{Anderle:2012rq}.)

In the current fragmentation region, the factorization formulas for the semi-inclusive structure functions, expanded to order $\alpha_s$ in the hard part, are given by
\begin{eqnarray}
2 F_1^h(\xbj,z_h,Q^2)
&=& \sum_{q ,\bar{q}} e_q^2\, 
    \bigg\{ f_q(\xbj,Q^2)\, D_q^h(z_h,Q^2)      \no
& & \hspace*{-2.2cm}
+\, \frac{\alpha_s(Q^2)}{2 \pi}
    \Big[ f_{q} \otimes C_{qq}^1 \otimes D_q^h
        + f_{q} \otimes C_{gq}^1 \otimes D_g^h
        + f_{g} \otimes C_{qg}^1 \otimes D_q^h
    \Big](\xbj,z_h,Q^2)
\bigg\} ,                                     \\
\frac{1}{x} F_L^h(\xbj,z_h,Q^2) 
&=& \sum_{q ,\bar{q}} e_q^2\,                   \no
& & \hspace*{-2.1cm}
    \times \frac{\alpha_s(Q^2)}{2 \pi}
    \Big[ f_{q} \otimes C_{qq}^L \otimes D_q^h 
        + f_{q} \otimes C_{gq}^L \otimes D_g^h
        + f_{g} \otimes C_{qg}^L \otimes D_q^h
    \Big](\xbj,z_h,Q^2),
\end{eqnarray}
where $f_i$ and $D_j^h$ label the PDF of flavor $i$ in the proton and parton $j$ $\to$ hadron $h$ FF, respectively.
The functions $C^{1}_{ij}$ ($C^{L}_{ij}$) are the lowest-order hard scattering coefficient functions for the $F_1$ ($F_L$) structure functions, and the symbol $\otimes$ denotes the convolution integral over longitudinal momentum fractions,
    $\left[A\otimes{B}\right](x) 
    \equiv \int_x^1\, (\diff{z}/z)\, A(z) B(x/z)$.
Explicit expressions for $C^{1}_{ij}$ ($C^{L}_{ij}$) are given, for example, in the appendices of Ref.~\cite{deFlorian:1997zj}.

In all of the expressions above, kinematical corrections from a nonzero target and final state hadron mass are neglected.
To focus attention on the current fragmentation region, in our analysis we impose the kinematic cuts $z_h > 0.2$ and $W^2 > 10$~GeV$^2$, and for the hard scale we choose $Q^2 > m_c^2$.
Therefore, the kinematical corrections may not be entirely negligible~\cite{Boglione:2019nwk} at the energies of some of the experiments, although for now we set aside a fuller account of their effect to a future dedicated analysis of mass corrections in SIDIS.

\bibliography{bibliography}

\end{document}